\documentclass[preprint,1p,number]{elsarticle}

\usepackage{multirow}
\usepackage{graphicx}
\usepackage{dcolumn}
\usepackage{bm}
\usepackage{url}                                       
\usepackage{subfigure}
\usepackage{epstopdf}

\journal{Physica A}

\begin{document}

\begin{frontmatter}

\title{Assessing the effectiveness of real-world network simplification}
\author{Neli Blagus\corref{coraut}}
\ead{neli.blagus@fri.uni-lj.si}
\author{Lovro \v Subelj}
\ead{lovro.subelj@fri.uni-lj.si}
\author{Marko Bajec}
\ead{marko.bajec@fri.uni-lj.si}
\address{University of Ljubljana, Faculty of Computer and Information Science, Ljubljana, Slovenia}
\cortext[coraut]{Corresponding author. Tel.: +386 1 476 81 86.}

\begin{abstract}
Many real-world networks are large, complex and thus hard to understand,
analyze or visualize. Data about networks are not always complete,
their structure may be hidden, or they may change quickly over time.
Therefore, understanding how an incomplete system differs from a complete
one is crucial. In this paper, we study the changes in networks submitted
to simplification processes (i.e., reduction in size). We simplify
$30$ real-world networks using six simplification methods and analyze
the similarity between the original and simplified networks based
on the preservation of several properties, for example, degree distribution,
clustering coefficient, betweenness centrality, density and degree
mixing. We propose an approach for assessing the effectiveness of
the simplification process to define the most appropriate size of
simplified networks and to determine the method that preserves the
most properties of original networks. The results reveal that the
type and size of original networks do not affect the changes in the
networks when submitted to simplification, whereas the size of simplified
networks does. Moreover, we investigate the performance of simplification
methods when the size of simplified networks is $10\%$ that of the
original networks. The findings show that sampling methods outperform
merging ones, particularly random node selection based on degree and
breadth-first sampling. 
\end{abstract}

\end{frontmatter}


\section{\label{sec:intro}Introduction}

Over the past decade, network analysis~\cite{NBW06,Newman09} has
proved to be a suitable tool for describing diverse systems, understanding
their structure and analyzing their properties. However, the evolution
of the Web and the capability of storing large amounts of data have
caused the size of networked systems and thus their complexity to
increase. The algorithms for analyzing and visualizing networks appear
impractical for addressing very large systems. Therefore, different
methods have been proposed for the simplification of complex networks.

Simplification is a process that reduces the size of a network by
decreasing the number of nodes and links. The procedure is derived
from graph theory (e.g., partitioning~\cite{FM91} and blockmodeling~\cite{DBF05})
and was initially developed for compression and efficient graph storage~\cite{DL98,AM01}.
With the increasing complexity of networks, simplification methods
also support clearer visualization~\cite{Rafiei05,HBFB08} and efficient
analysis~\cite{LF06,LKJ06}. In addition to these benefits, analyzing
the changes undergone by networks under the effects of the simplification
process enables us to explore and explain the differences between
complete (i.e., original) and incomplete (i.e., simplified) systems
(e.g., when only partial insight into the structure of network is
available). 

Recently, network simplification has been extensively investigated
from different perspectives. Some studies have concentrated on the
simplification of specific networks, such as simplifying social networks
based on stability and retention~\cite{KHSA10}, sampling scale-free~\cite{SWM05}
or directed networks~\cite{ACBFGP12}, estimating different properties
under social network crawling~\cite{DB13}, sampling large dynamic
peer-to-peer networks with random walks~\cite{SRDSW09} or simplifying
flow networks by removing useless links~\cite{BBV00}. Other studies
have attempted to provide a sufficient fit to original networks and
thus observe the changes in network properties under the effects of
simplification, such as preserving the clustering coefficient~\cite{SC12},
degree distribution~\cite{SW05}, community structure~\cite{MP12},
spectral properties~\cite{GR07} or network connectivity~\cite{ZMT10}.

However, only a few studies have focused on comparing simplification
methods and measuring their success. Leskovec~et~al.~\cite{LF06}
observed properties of original and simplified networks submitted
to several simplification methods and measured their success based
on random walk similarity. Lee~et~al.~\cite{LKJ06} analyzed basic
network properties under the effects of three simplification methods
and revealed characteristic patterns of changes in properties. H{\"u}bler~et~al.~\cite{HKBG08}
compared their simplification algorithm to existing ones by measuring
the average distance of properties between original and simplified
networks. Toivonen~et~al.~\cite{TZHH11} studied the compression
of weighted networks and measured the method's efficiency according
to the running time and cost of the compressed network representation.
Doer and Blenn~\cite{DB13} tested the convergence of different
properties under three traversal algorithms applied to a single large
social network. The findings of the aforementioned analyses indicate
that the performances of simplification methods vary; however, the
common weakness of these studies is the small set of networks considered.

Despite the above-described efforts, several open questions remain
concerning the simplification of complex networks, such as those regarding
(Q1) how to evaluate the similarity between original and simplified
network, (Q2) how small simplified networks should be and ultimately
(Q3) what simplification method should be used. In this paper, we
address these questions and propose an approach for assessing the
effectiveness of the simplification process. We analyze $30$ real-world
networks of different size and origin under the effects of six different
simplification methods. We compare the original and simplified networks
based on several network properties (e.g., degree distribution, clustering
coefficient~\cite{WS98}, betweenness centrality~\cite{Brandes01},
degree mixing~\cite{Newman02} and transitivity~\cite{wasserman94})
(Q1). The selection of these properties is supported by their common
use in similar studies~\cite{LF06,LKJ06}. Moreover, we propose
a measure for determining the most appropriate size of simplified
networks for preserving the observed properties (Q2) and for determining
under which method the simplified networks fit the original ones most
closely (Q3). We also study the impact of the original network size and
type on the effectiveness of the simplification process.

The rest of the paper is structured as follows. Section~\ref{sec:methods}
focuses on the simplification methods and real-world networks used
in the study and describes the proposed measure. In section~\ref{sec:analys},
we report and formally discuss the results of the analysis. Finally,
section~\ref{sec:conc} concludes the paper and suggests directions
for future research.


\section{\label{sec:methods}Methods and data}

\subsection{Simplification methods}

\begin{figure}[t]
        \centering
        \subfigure[\label{subfig:RN}]{\includegraphics[width=0.23\columnwidth]{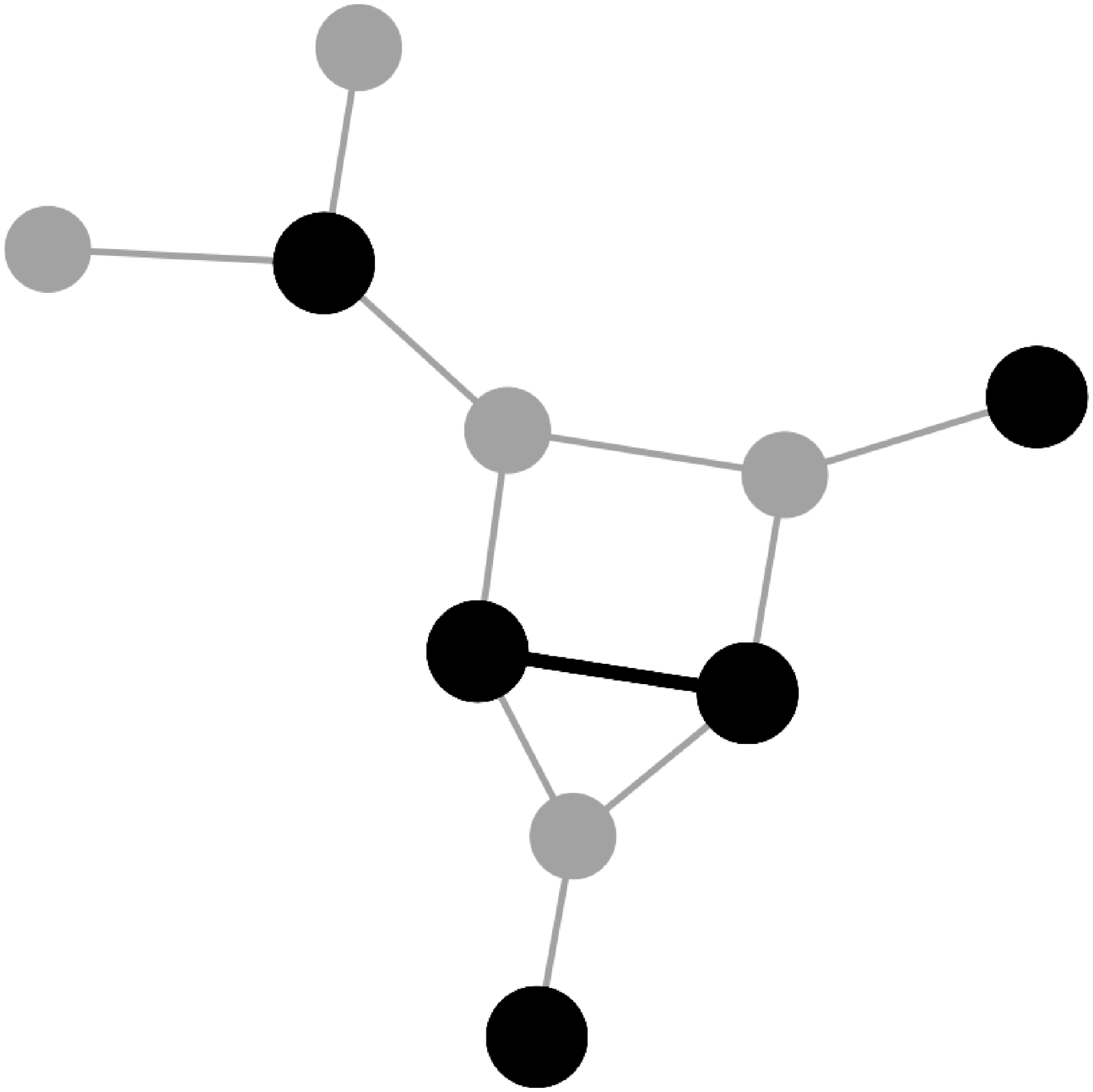}}\quad
				\subfigure[\label{subfig:RD}]{\includegraphics[width=0.23\columnwidth]{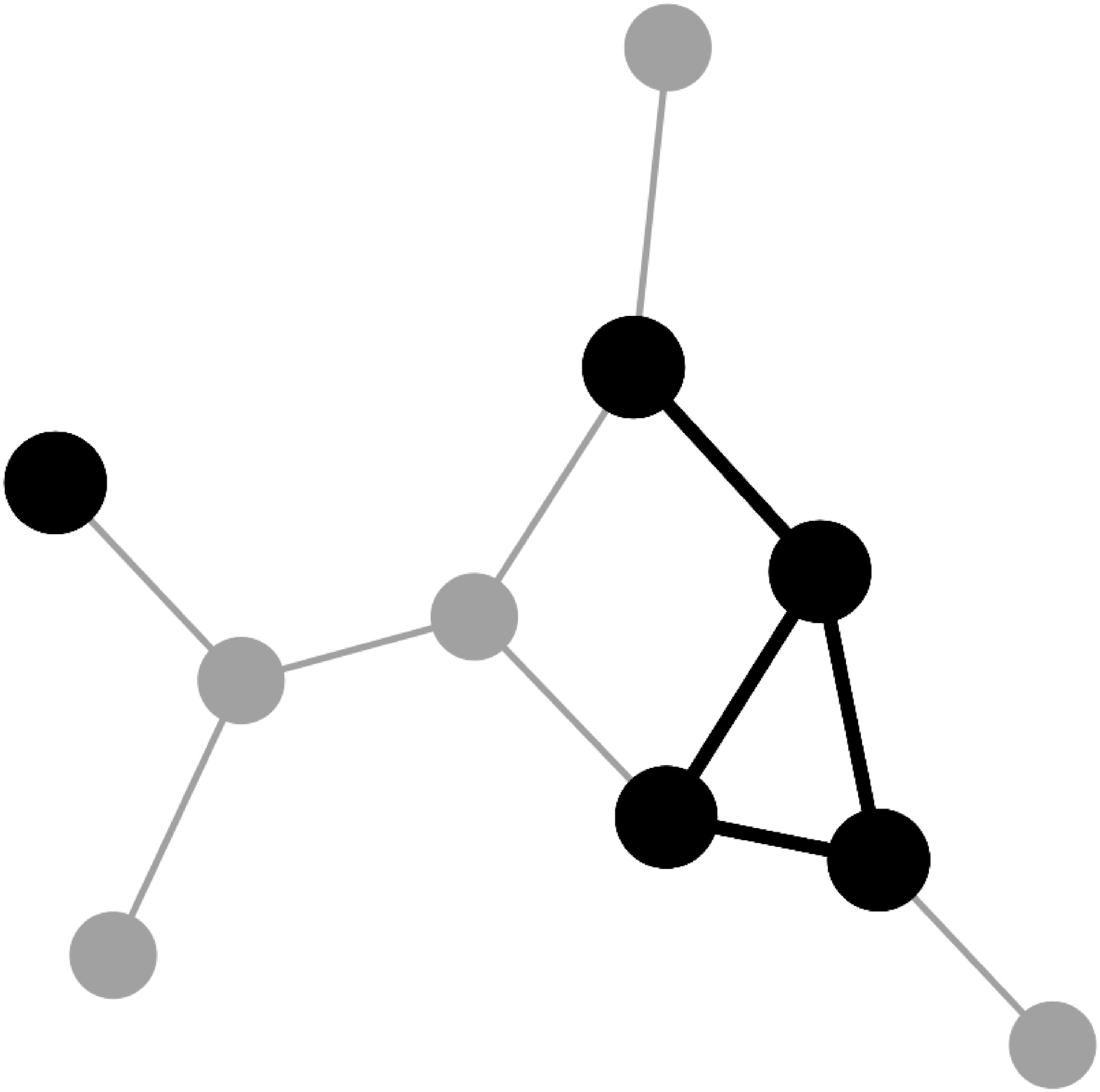}}\quad
				\subfigure[\label{subfig:RL}]{\includegraphics[width=0.23\columnwidth]{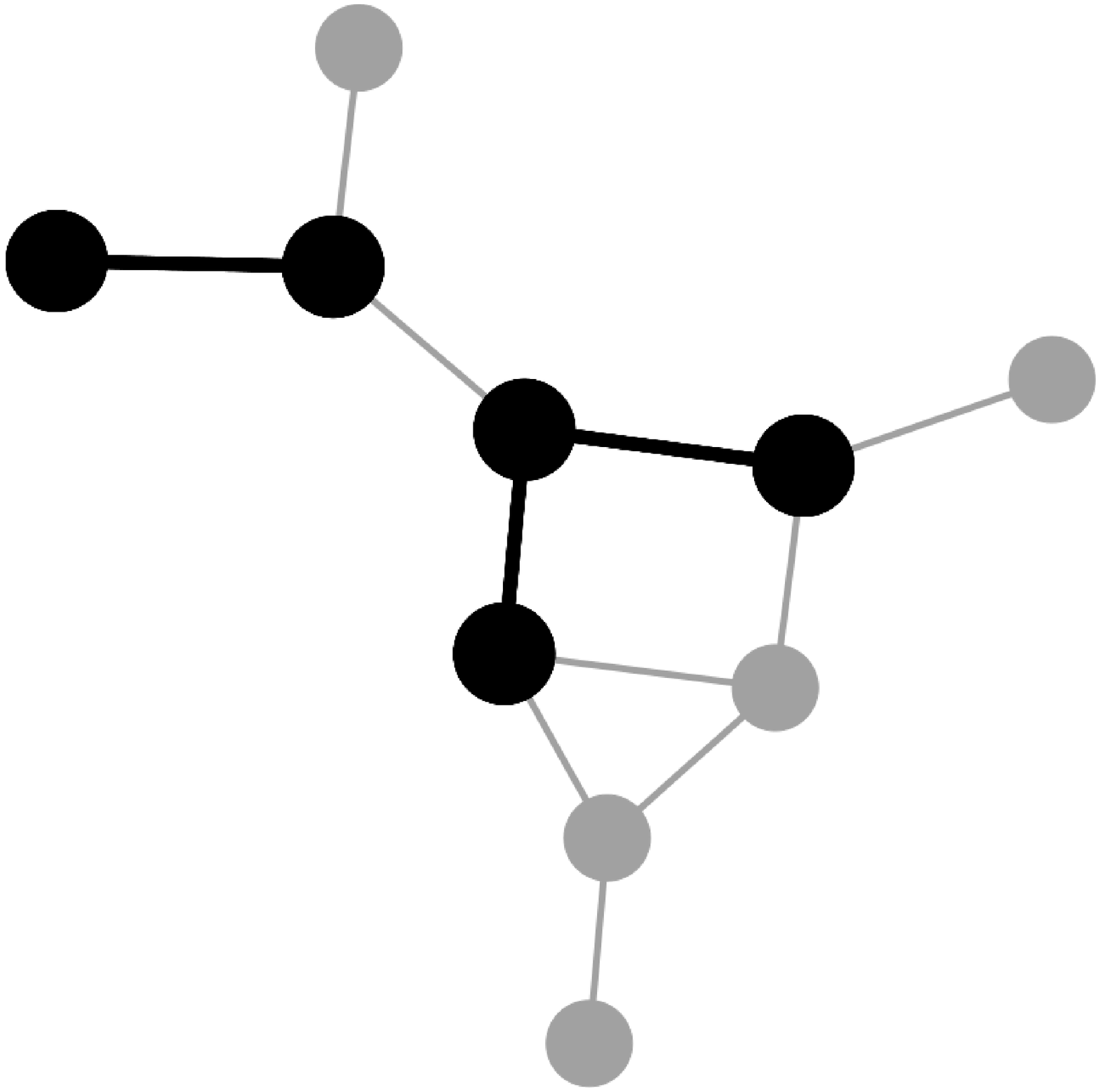}}\quad
				\subfigure[\label{subfig:SB}]{\includegraphics[width=0.23\columnwidth]{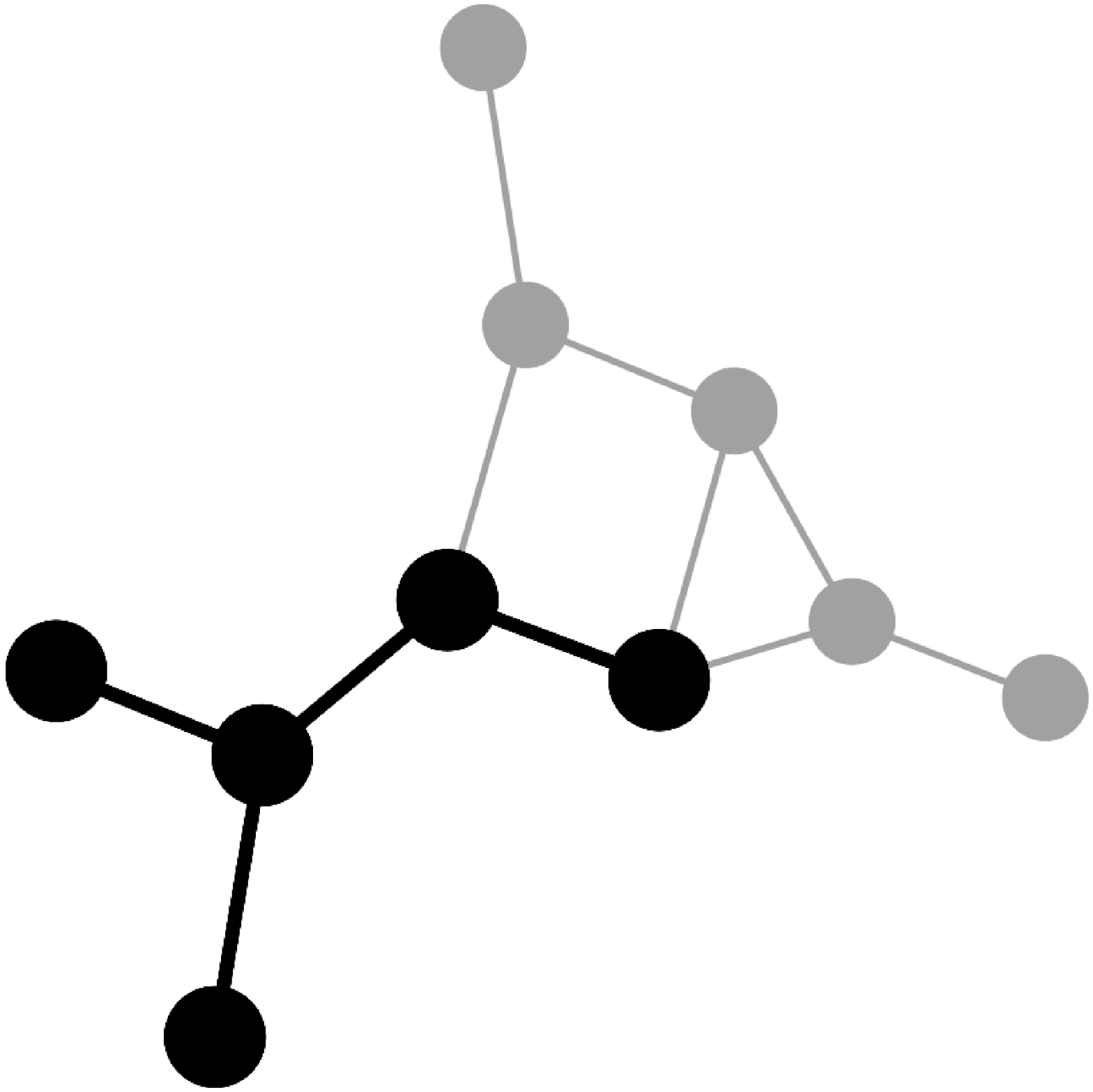}}
\caption{Sampling methods applied to a small sample network for $s=0.5$. Black
nodes represent simplified networks obtained with~\subref{subfig:RN}
selecting nodes uniformly at random (RN),~\subref{subfig:RD} selecting
nodes with probability proportional to degree (RD),~\subref{subfig:RL}
selecting links uniformly at random (RL) and~\subref{subfig:SB}
performing breadth-first search starting at a randomly selected node
(BF). In the last method, BF ensures a connected network, whereas
in other methods, this is not always the case.}
\label{fig:sampl} 
\end{figure}

Several authors have proposed a broad collection of simplification
methods, which can be divided into two general classes. Those in the
first class are sampling methods in which a simplified network is
represented by a random sample of the original network (e.g., random
node selection~\cite{KFCLCP05}, random link selection~\cite{ANK11},
snowball sampling~\cite{IF11}, random walk sampling~\cite{LF06}
and forest fire~\cite{LF06}). Methods in the second class obtain
simplified networks by merging nodes and links into supernodes and
superlinks based on different characteristics, such as the distance
between nodes (e.g., cluster-growing and box-tiling renormalization~\cite{GSM07}),
node and link attributes (e.g., link weights~\cite{TZHH12} and
node attributes~\cite{ZCY09}) or community structure (e.g., balanced
propagation and modularity optimization~\cite{BSB12}).

In this study, we adopt four basic sampling methods (Fig.~\ref{fig:sampl}).
Random node~\cite{KFCLCP05} (RN) and random link selection~\cite{ANK11}
(RL) create sampled networks with nodes or links selected uniformly
at random. Simplified networks under random node selection based on
degree~\cite{LF06} (RD) consist of randomly selected nodes, where
the probability of selecting a node is proportional to the node's
degree. In breadth-first sampling (BF), a random node with its broad
neighborhood is selected into the sample using the breadth-first search
strategy. The main advantages of these methods are simplicity, and
thus efficient implementation with low time complexity, and adjustability,
which enables setting the size of the simplified network in advance.

\begin{figure}[t]
        \centering
        \subfigure[\label{subfig:BP}]{\includegraphics[width=0.23\columnwidth]{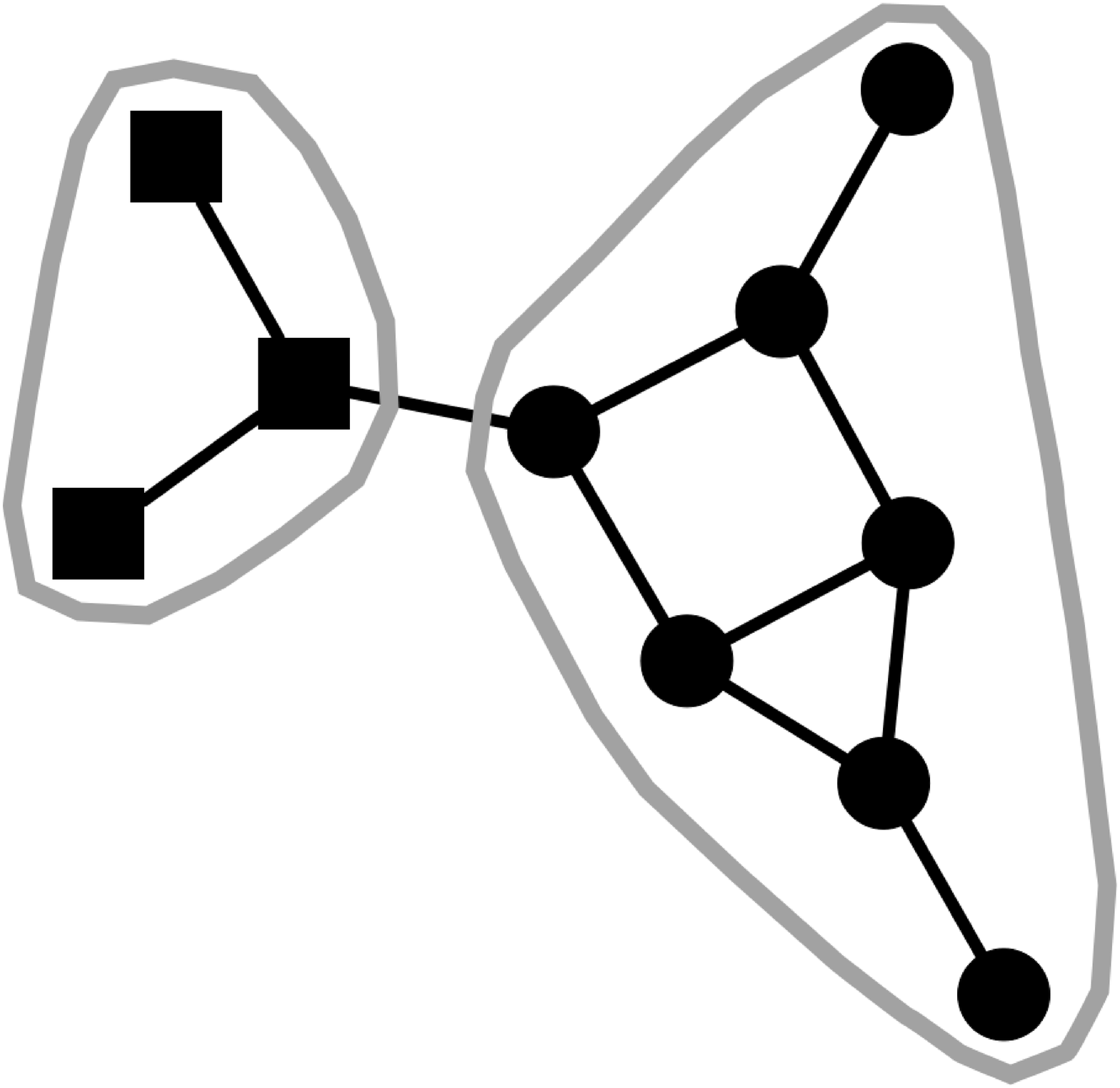}}\quad
				\subfigure[\label{subfig:CG}]{\includegraphics[width=0.23\columnwidth]{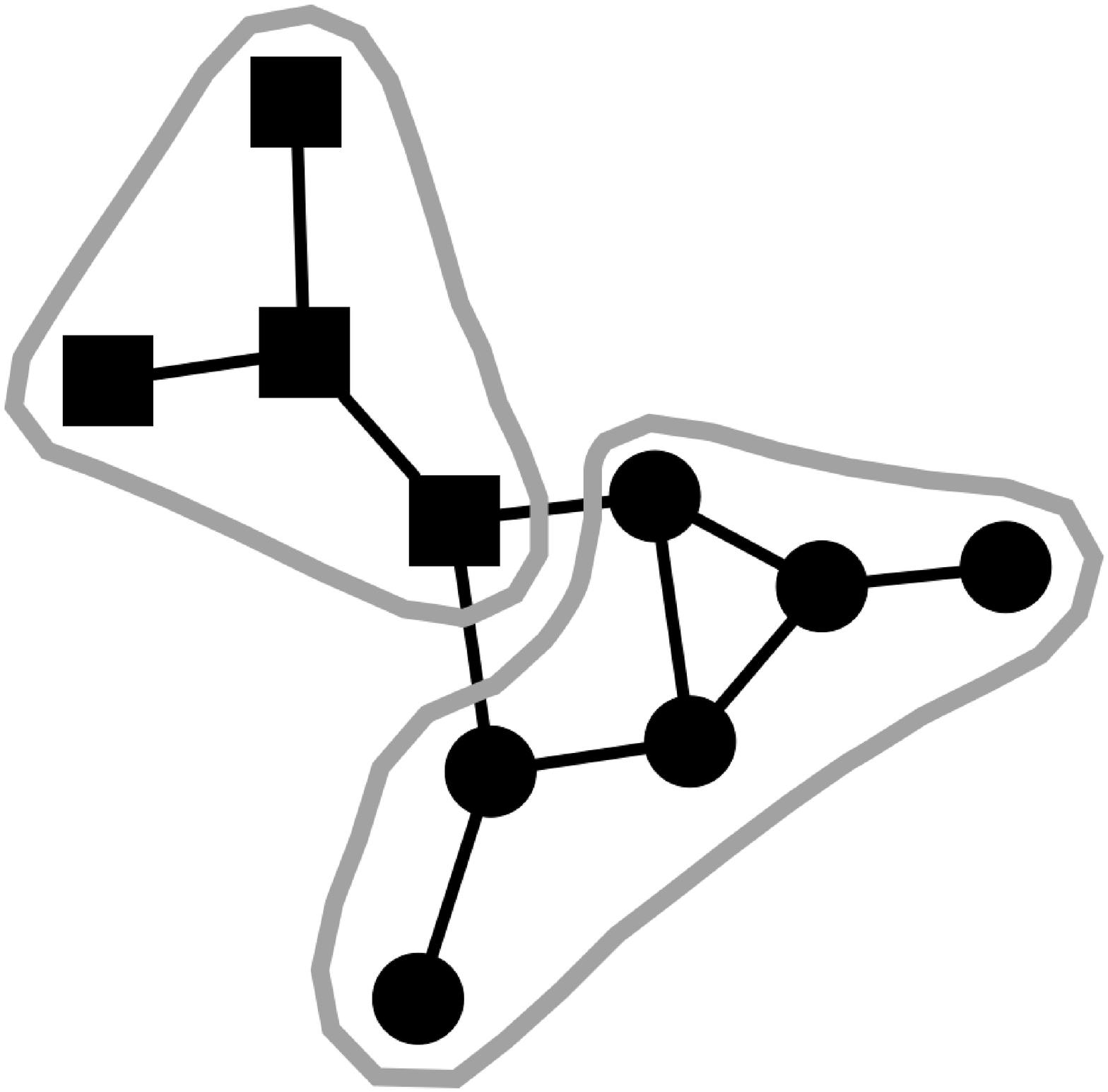}}\quad
				\subfigure[\label{subfig:sn}]{\includegraphics[width=0.23\columnwidth]{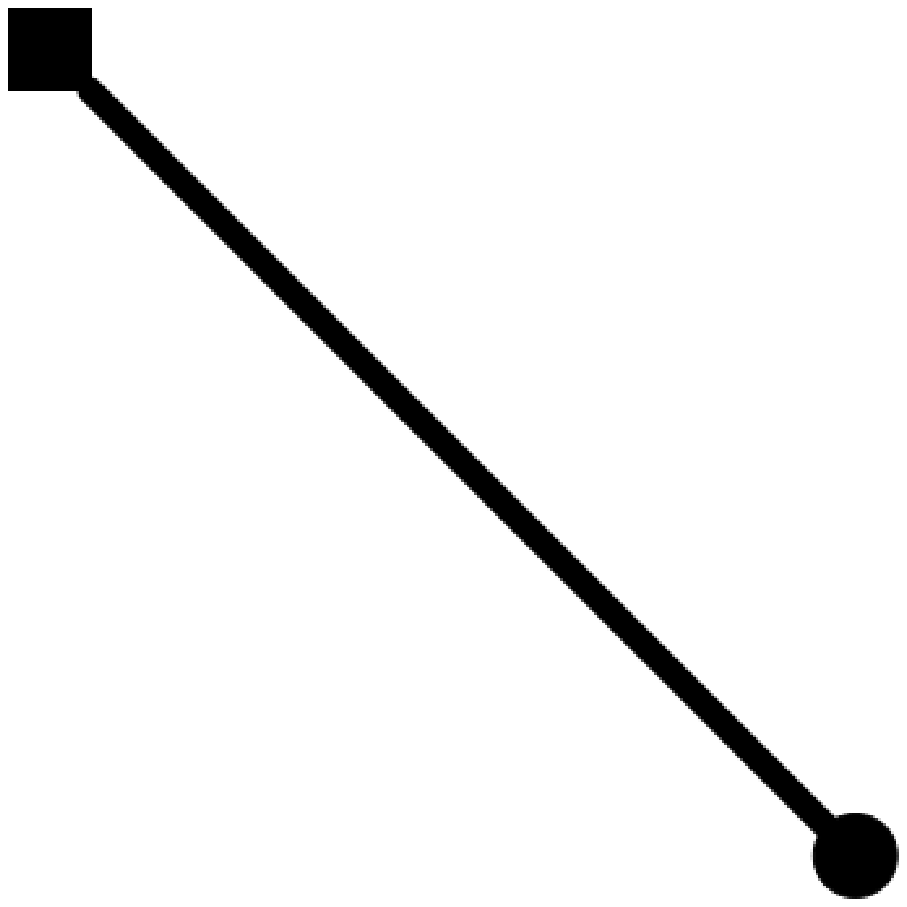}}
\caption{Merging methods applied to a small sample network. The shape of the
nodes indicates~\subref{subfig:BP} the nodes' community membership
(BP) and~\subref{subfig:CG} whether the nodes are at a distance
less than $5$ ($c=2$) within one box (CG). Communities and boxes
are marked by a gray contour. The simplified network (shown for both
cases in~\subref{subfig:sn}) is obtained by merging nodes inside
one community or box into supernodes.}

\label{fig:merg} 
\end{figure}

Sampling methods outperform merging ones in terms of the advantages
listed above. Still, we consider two methods from the merging class
(Fig.~\ref{fig:merg}). We use merging nodes based on community detection,
where supernodes are identified by communities revealed by balanced
propagation~\cite{Subelj11} (BP). We also employ cluster-growing
renormalization~\cite{GSM07} (CG), which incrementally grows supernodes
from randomly selected seed nodes within a distance not larger than
$c$ (the nodes within one supernode are at most $2\cdot c+1$ steps
apart). Both methods proved well in analyzing the invariance of network
density under different renormalizations~\cite{BSB12}.

We define $s$ as the number of nodes in the simplified network, measured
as the fraction of nodes in the original network. For sampling, we
set the sizes of the simplified networks as varying from $1\%$ to
$50\%$ of the original networks ($s=0.01$ and $s=0.05$-$0.50$
with a step size of $0.05$). For BP, we set the parameters of the
algorithm as suggested in~\cite{Subelj11}. With CG, we cannot control
the size of the simplified network; still, we can change the distance
between the nodes within one supernode. Therefore, the parameter $c$
ranges from two to six, where smaller values indicate a smaller number
of nodes within one supernode and thus a larger simplified network.

\subsection{Network data}

A diverse set of real-world systems is analyzed. We consider $30$
networks of different origin (e.g., information, technological and
social) and size (varying from a few thousand to a few hundred thousand nodes),
listed in Table~\ref{tbl:nets}. Due to the large number of networks
considered, a detailed description is omitted here.

For BP, CG and BF, all networks are considered to be undirected, although
some of them are directed. To avoid comparing networks of different
complexity, we remove self-loops and multiple links from all networks
for simplification via merging methods.

\begin{table}[t]
\scriptsize
\centering
\caption{\label{tbl:nets}Real-world networks ($n$ and $m$ correspond to the number of nodes and links, respectively).}
\begin{tabular}{lcrr}
\hline\noalign{\smallskip}
Network & Type & \multicolumn{1}{c}{$n$} & \multicolumn{1}{c}{$m$}\\\noalign{\smallskip}\hline\noalign{\smallskip}
\textit{High E. Particle Phys.}~\cite{KDD} & \multirow{4}{*}{Citation} &	$27240$ &	$342437$ \\
\textit{High E. Phys.}~\cite{LKF05} & &	$34546$ &	$421578$ \\
\textit{NBER US} patents~\cite{HJT01} &	&	$240548$ & $561060$ \\
\textit{Citeseer} publications~\cite{BLG98} &	&	$384413$ &	$1764929$ \\
\noalign{\smallskip}
\textit{PGP} web-of-trust~\cite{BPSGA04}	&	\multirow{5}{*}{Collaboration}	&	$10680$ &	$24340$ \\
\textit{High E. Phys.} archive~\cite{LKF07} & &	$12008$ &	$237010$ \\
\textit{Astro Phys.} archive~\cite{LKF07} & &	$18772$ &	$396160$ \\
\textit{Cond. Matters} archive~\cite{LKF07} & &	$23133$ &	$186936$ \\
Computer science~\cite{YL12} & &	$317080$ &	$1049866$ \\
\noalign{\smallskip}
\textit{Digg} user reply~\cite{DSJS09} & \multirow{4}{*}{Communication} &	$30398$ &	$87627$ \\
Emails at \textit{Enron}~\cite{Yl12a}	& &	$36692$ &	$367662$ \\
\textit{Facebook} wall post~\cite{VMCG09} & &	$46952$ &	$876993$ \\
Emails at EU res. inst.~\cite{LKF07} & &	$265214$ &	$420045$ \\
\noalign{\smallskip}
\textit{Amazon} products $1$~\cite{Yl12a} & \multirow{2}{*}{Co-purchase}	&	$334863$ &	$925872$ \\
\textit{Amazon} products $2$~\cite{LAH07} & &	$403394$ &	$3387388$ \\
\noalign{\smallskip}
\textit{Flickr} images metadata~\cite{ML12} & \multirow{1}{*}{Co-occurence}	&	$105938$ &	$2316948$ \\
\noalign{\smallskip}
\textit{Oregon} aut. systems~\cite{newmanData} & \multirow{3}{*}{Internet}	&	$22963$ &	$48436$ \\
\textit{Gnutella} file sharing $1$~\cite{LKF07}	&	&	$36682$ &	$88328$ \\
\textit{Gnutella} file sharing $2$~\cite{LKF07}	&	&	$62586$ &	$147829$ \\
\noalign{\smallskip}
\textit{Foldoc} dictionary~\cite{BMZ02}	& \multirow{1}{*}{Information}	&	$13356$ &	$120238$ \\
\noalign{\smallskip}
\textit{Wikipedia} votes~\cite{LHK10}	&	\multirow{6}{*}{On-line social} &	$7115$ &	$103689$ \\
\textit{Brightkite} friendship~\cite{CML11} &	&	$58228$ &	$214078$ \\
\textit{Epinions} trust~\cite{RAD03} & &	$75879$ &	$508837$ \\
\textit{Slashdot} friendship~\cite{LLDM09} &	&	$82168$ &	$948464$ \\
\textit{Wikipedia} interactions~\cite{MAC11} &	&	$186485$ &	$740397$ \\
\textit{Gowalla} friendship~\cite{CML11} &	&	$196591$ &	$1900654$ \\
\noalign{\smallskip}
Broad-topic queries~\cite{kleinberg99}	& \multirow{4}{*}{Web graph}	&	$6175$ &	$16150$ \\
\textit{google.com} internal~\cite{PFPDV07} &	&	$15763$ &	$171206$ \\
\textit{nd.edu} domain~\cite{AJB99} &	&	$325729$ &	$1497134$ \\
\textit{Baidu} articles~\cite{NSWRQY11} &	&	$415641$ &	$3284387$ \\
\noalign{\smallskip}\hline
\end{tabular}
\end{table}

\subsection{Assessment approach}

To perform a fair and sound assessment, we first address the aforementioned
questions concerning the comparison approach (Q1) and the size to
which a certain network should be simplified (Q2). To address Q1,
we select a set of local and global network properties to be observed.
To address Q2, we introduce a simple measure that takes into account
all of the selected properties and for each network calculates the
simplified size that would best preserve the observed properties.
The specific size of the simplified networks is then used in a further
analysis to compare the selected simplification methods (Q3).

\subsubsection{\label{subsec:Q1}Comparing original and simplified networks}

We compare networks based on eight fundamental global and local properties.
The global properties are expressed by a single value for each network
and include density (the ratio of existing links to all possible links),
degree mixing (the tendency of nodes connecting to similar ones~\cite{Newman02})
and transitivity (the number of closed triplets over the total number
of triplets~\cite{wasserman94}). The local properties are described
by a distribution for all nodes in the network and comprise degree,
in-degree and out-degree (the number of neighbors of each node), local
clustering coefficient (the proportion of connected neighbors of each
node~\cite{WS98}) and betweenness centrality (the number of shortest
paths between all nodes going through each node~\cite{Brandes01}).

For comparison, we define two similarity measures, one based on the
selected global properties and one on the selected local properties.
The global similarity measure is used to determine how correlated
the global properties in the observed original networks and their
simplified version are. The correlation is measured with Spearman’s
correlation coefficient $\rho$. $\rho$ indicates the extent to which
one variable decreases as another increases. In our analysis, we calculate
$\rho$ for each selected simplification method and each size of the
simplified networks for all networks together.

The comparison based on the selected local properties is expressed
using the Kolmogorov-Smirnov $D$-statistic (Kolmogorov-Smirnov test
checks the null-hypothesis, i.e., that the distributions of two properties
are the same; the D-statistic measures the distance between the observed
distributions). The $D$-statistic for each network and its simplified
version is calculated for each simplification method separately. The
values for comparison based on $\rho$ and the $D$-statistics are
averaged over ten simplifications of each network, each simplification
method and each size of the simplified networks.

The selection of properties and their relevance in assessing the effectiveness
of network simplification greatly depends on the purpose of the simplification
being performed. The selection of particular properties in this analysis
is only supported by their common use in similar studies (e.g.,~\cite{LF06,LKJ06})
and serve to demonstrate the effectiveness of the proposed approach.
Note that comparing networks based on other sets of properties may
lead to different results.

In the literature, we can find studies that have performed similar
comparisons to a limited extent. In ~\cite{SWM05} the authors proved
that RN does not preserve the degree distribution of scale-free networks.
Moreover, RN and RL sampling are biased toward nodes with high degrees,
which affects the degree distribution~\cite{LF06}. However, Lee
et al.~\cite{LKJ06} proved that RN and RL overestimate the degree
and betweenness centrality exponent, whereas both methods retain the
assortativity of original networks. Merging methods decreases the
density~\cite{BSB12}, but the relationship between density and
network size remains invariant after simplification.

\subsubsection{Determining simplified network sizes}

To determine the size to which a specific network can be decreased
while preserving most of the observed properties, the following approach
is used. For each simplification method and each global and local
property, we rank sizes with respect to $\rho$ and the $D$-statistic,
respectively. The network size that best fits a specific property
receives rank 0, the next best one receives rank 1 and so on. Next,
we sum the ranks for each size and divide the sum by the greatest
possible sum of ranks to normalize the result to the interval $[0,1]$.
Thus, the measure $A$ is defined as 
\begin{equation}
	A = \frac{1}{(n_s - 1) \cdot n_p} \sum_{i=1}^{n_p} r_i,
\end{equation}
where $n_{s}$ denotes the number of different sizes, $n_{p}$ denotes
the number of properties, $i$ indexes the properties (the order is
not important) and $r_{i}$ is the rank of the $i-$th property. $A$
is thus the normalized total rank assigned to a specific size by the
observed properties.

Table~\ref{tbl:exampleQ2} shows an example of the measure A calculated
by comparing six different sizes for a simplified network, taking
into account the measure $\rho$ a specific size receives for each
of the three observed global properties. In this example, the most
appropriate size for preserving global properties is S$6$.

\begin{table}[t]
\scriptsize
\centering
\caption{\label{tbl:exampleQ2}An illustrative example of the assessment approach. (left) 
The results of a comparison between simplified and original networks based on global properties. 
(right) The results after ranking sizes for each property. P\textit{i} denotes properties 
and S\textit{i} sizes of simplified networks.}
	\begin{tabular}{lccc}
		\hline\noalign{\smallskip}
		 & P1 & P2 & P3 \\ 
		\hline\noalign{\smallskip}
		S1 \hspace{0.1cm} \vline & $0.84$    & $0.69$    & $0.75$\\
    S2 \hspace{0.1cm} \vline & $0.88$    & $0.89$    & $0.87$\\
    S3 \hspace{0.1cm} \vline & $0.90$    & $0.92$    & $0.89$\\
    S4 \hspace{0.1cm} \vline & $0.96$    & $0.95$    & $0.88$\\
    S5 \hspace{0.1cm} \vline & $0.91$    & $0.94$    & $0.92$\\
    S6 \hspace{0.1cm} \vline & $0.93$    & $0.96$    & $0.90$\\
		\hline\noalign{\smallskip}
	\end{tabular}%
	\hspace{1cm}
  \begin{tabular}{lccccc}
		\hline\noalign{\smallskip}
		 & P1 & P2 & P3 \hspace{0.1cm} & Sum & $A$\\
		\hline\noalign{\smallskip}
		S1 \hspace{0.1cm} \vline & $5$    & $5$    & $5$ \hspace{0.1cm} \vline & $15$ & $1.000$ \\
    S2 \hspace{0.1cm} \vline & $4$    & $4$    & $4$ \hspace{0.1cm} \vline & $12$ & $0.800$ \\
    S3 \hspace{0.1cm} \vline & $3$    & $3$    & $2$ \hspace{0.1cm} \vline & $8$ & $0.533$ \\
    S4 \hspace{0.1cm} \vline & $0$    & $1$    & $3$ \hspace{0.1cm} \vline & $4$ & $0.267$ \\
    S5 \hspace{0.1cm} \vline & $2$    & $2$    & $0$ \hspace{0.1cm} \vline & $4$ & $0.267$ \\
    S6 \hspace{0.1cm} \vline & $1$    & $0$    & $1$ \hspace{0.1cm} \vline & $2$ & $0.133$ \\
		\hline\noalign{\smallskip}
	\end{tabular}%
\end{table}

\subsubsection{Comparing simplification methods}

Finally, we compare the different methods for a given size of a simplified
network. We rank the methods and measure their effectiveness using
a modified version of the measure $A$ described in the previous subsection:
\begin{equation}
	A = \frac{1}{(n_m - 1) \cdot n_p} \sum_{i=1}^{n_p} r_i,
\end{equation}
where $n_{m}$ is the number of different methods.

With the described measure, we regard all properties as equally important.
Still, depending on the purpose of the simplified networks considered
and the method by which those networks are analyzed, one property
can be more essential than another. With respect to importance, we
can assign weights $w$ to the properties; thus, the measure $A$
becomes 
\begin{equation}
	A_w = \frac{1}{n_m \cdot \sum_{i=1}^{n_p} w_i} \sum_{i=1}^{n_p} r_i \cdot w_i,
\end{equation}
where $w$ is the vector of weights and thus $w_{i}$ denotes the
weight of property $i$.

For simplicity, we omit the analysis performed based on the measure
$A_{w}$ and thus assume all properties are equally important.


\section{\label{sec:analys}Analysis and discussion}

The analysis consists of two stages. First, we determine the size
of the simplified networks that ensure adequate preservation of the
observed properties. Second, we compare the effectiveness of different
methods for a specific size of the simplified networks.

\subsection{\label{subsec:effect}Effectiveness of the simplification process with 
respect to the size of the simplified networks}
\begin{figure}[t]
        \centering
        \subfigure[\label{subfig:freqsampl}]{\includegraphics[width=0.47\columnwidth]{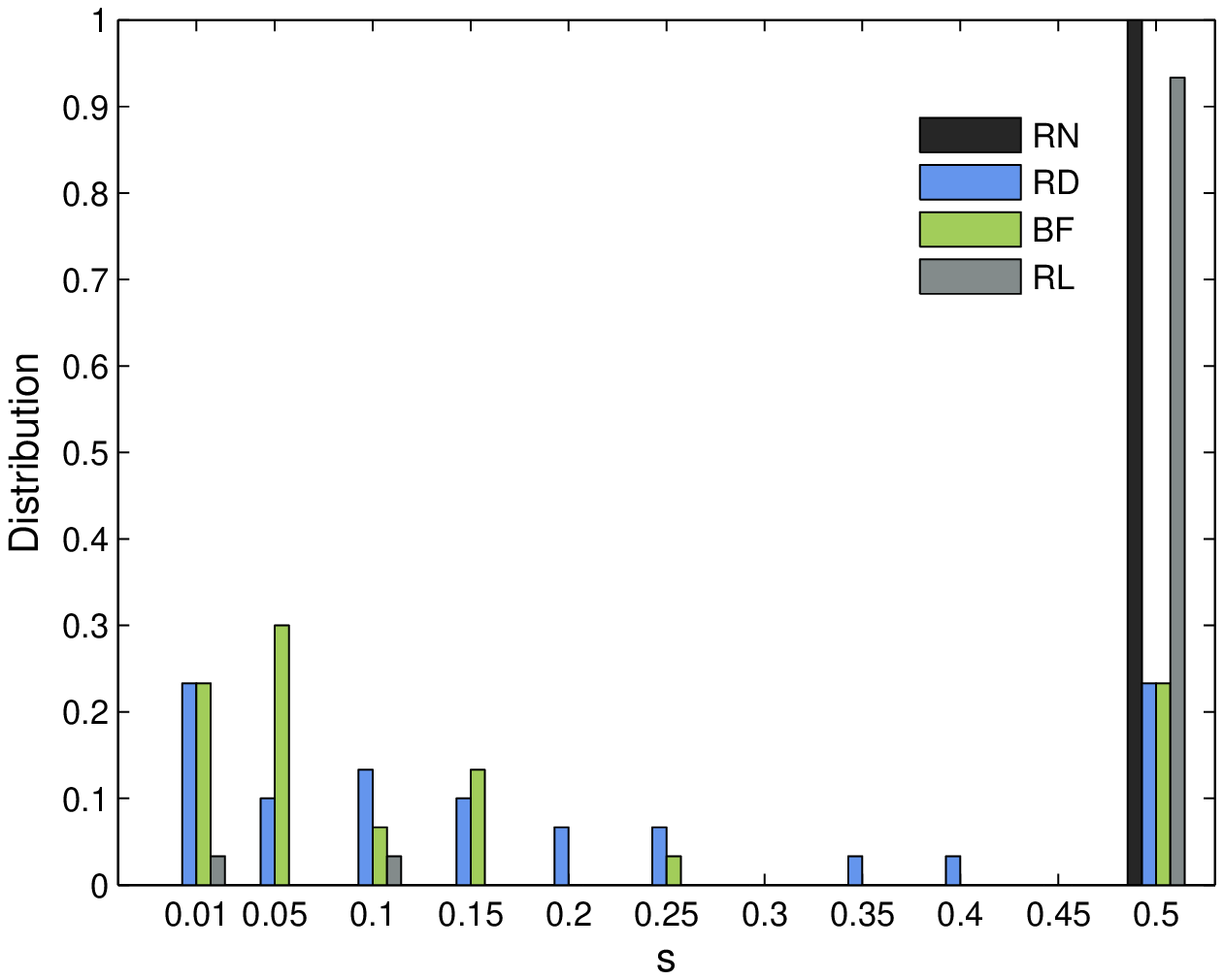}}\quad
								\subfigure[\label{subfig:localsampl}]{\includegraphics[width=0.47\columnwidth]{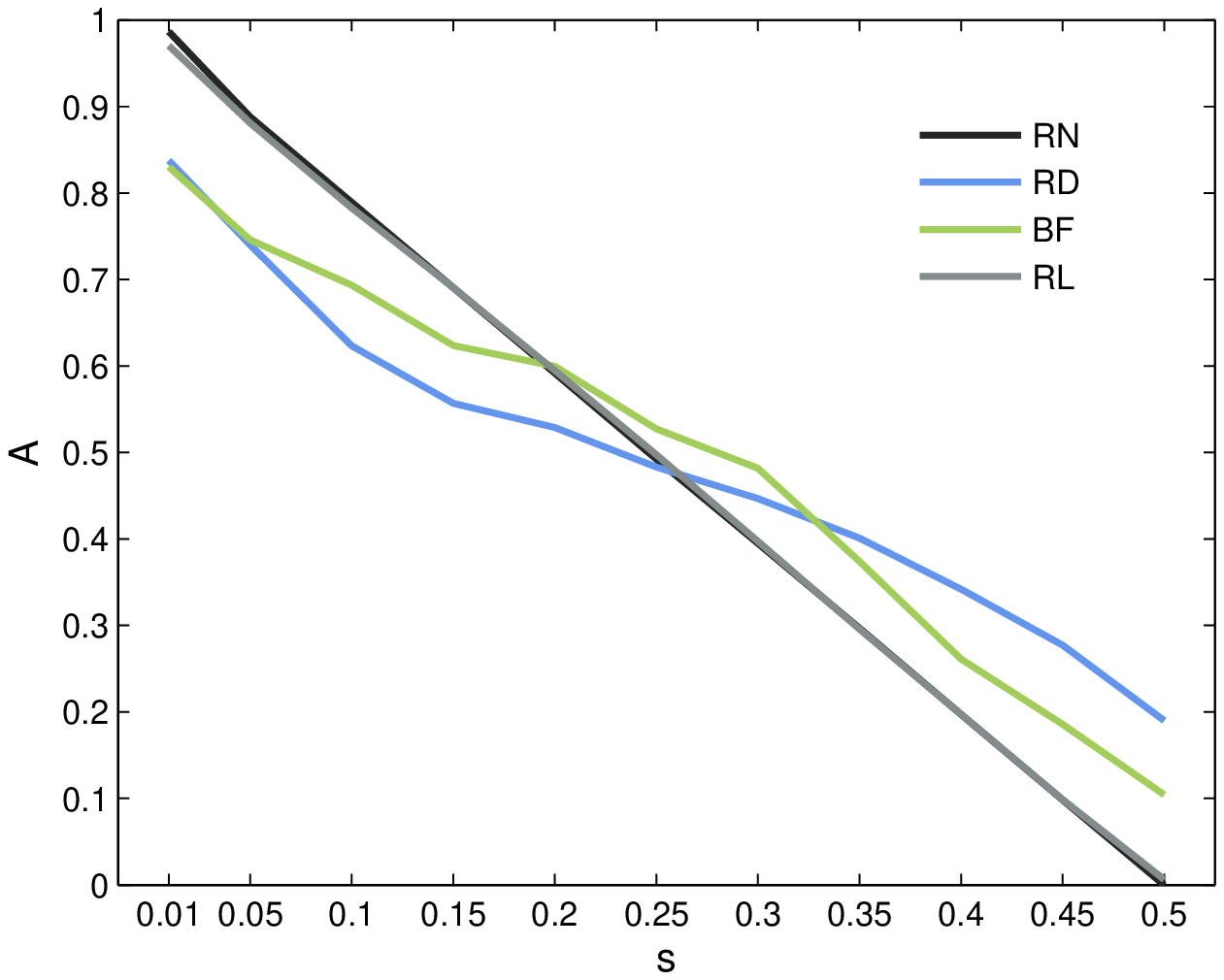}}\quad
				\subfigure[\label{subfig:globsampl}]{\includegraphics[width=0.47\columnwidth]{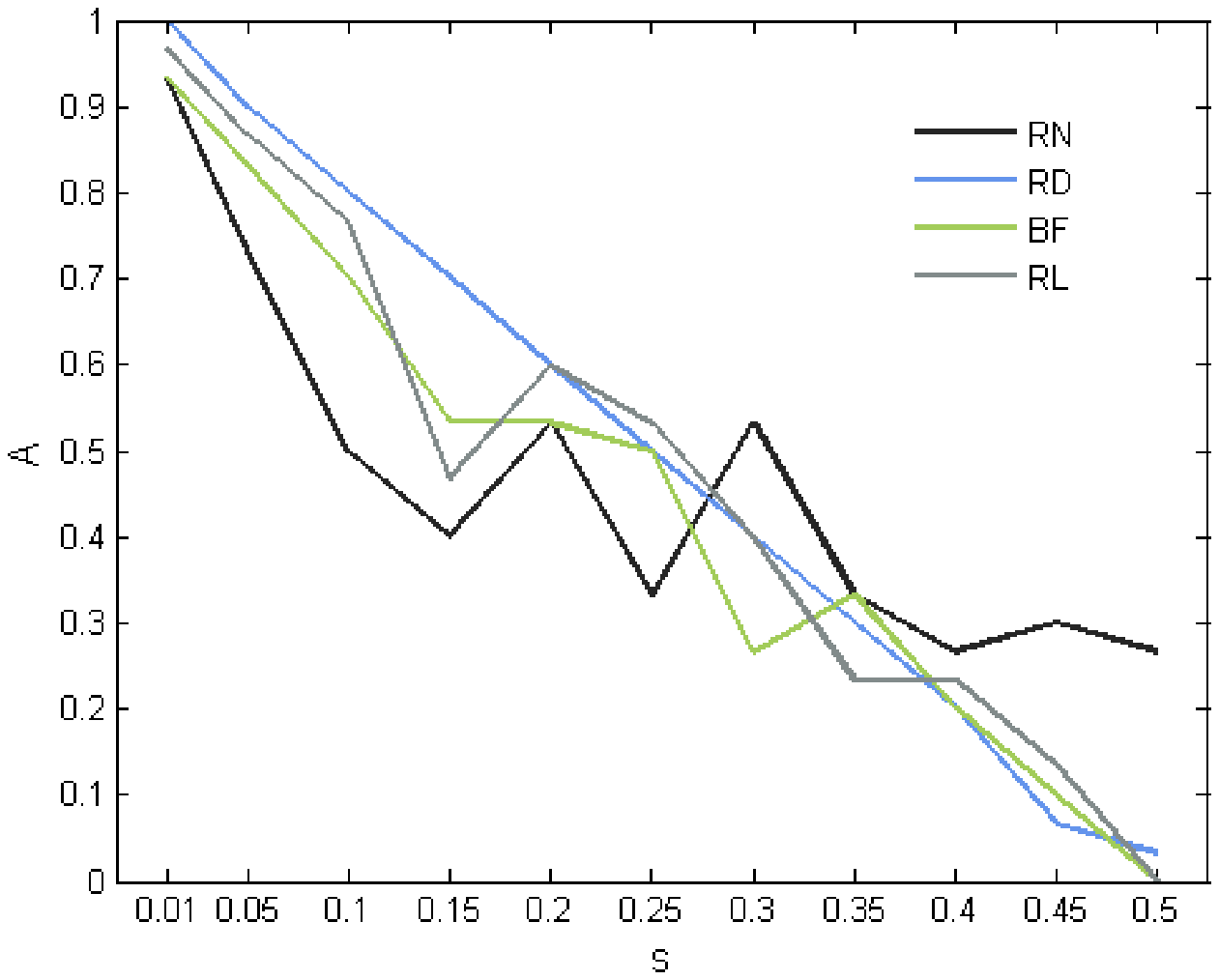}}
        \caption{The results for the sampling methods. \subref{subfig:freqsampl} Portion of networks with the best size 
				equal to $s$. \subref{subfig:localsampl} Distance between the original and simplified networks (average $A$ over 
				all networks) based on the local properties. \subref{subfig:globsampl} Distance between original and simplified 
				networks based on the global properties. }\label{fig:bestsizesampl}
\end{figure}

First, we analyze the effect of simplified network size on the effectiveness
of the simplification process. As expected, the results reveal that
in the majority of cases, the largest simplified networks ($c=2$
for CG and $s=0.5$ for sampling methods) are more similar to the
original networks and thus better fit the original networks' properties.
However, the main goal of the simplification is to sufficiently reduce
large networks to allow for easier analysis and understanding, which
is achieved when the simplified networks are smaller. Therefore, we
define the best size as the local minimum of $A$ achieved at the
smallest simplified network size (we assume that $A=1$ for $s=0$ and 
take the global minimum if it is also local).

\begin{table}[t]
\scriptsize
\centering
\caption{\label{tbl:resS1}The best sizes $c$ or $s$ for the preservation of the local properties with corresponding $A$.}
\begin{tabular}{lccc}
\hline\noalign{\smallskip}
Network & CG & RD & BF \\\noalign{\smallskip}\hline\noalign{\smallskip}
\textit{High E. Particle Phys.} & $2$ $(0.00)$ & $0.20$ $(0.18)$ & $0.05$ $(0.38)$ \\
\textit{High E. Phys.} & $2$ $(0.08)$ & $0.25$ $(0.14)$ & $0.10$ $(0.44)$ \\
\textit{NBER US} patents & $2$ $(0.25)$ & $0.35$ $(0.20)$ & $0.01$ $(0.44)$ \\
\textit{Citeseer} publications &$2$ $(0.00)$ & $0.20$ $(0.08)$ & $0.01$ $(0.58)$ \\
\noalign{\smallskip}
\textit{PGP} web-of-trust	 & $4$ $(0.33)$ & $0.25$ $(0.40)$ & $0.05$ $(0.77)$ \\
\textit{High E. Phys.} archive & $2$ $(0.00)$ & $0.05$ $(0.73)$ & $0.05$ $(0.83)$ \\
\textit{Astro Phys.} archive & $2$ $(0.00)$ & $0.50$ $(0.17)$ & $0.05$ $(0.50)$ \\
\textit{Cond. Matters} archive & $2$ $(0.00)$ & $0.50$ $(0.13)$ & $0.05$ $(0.70)$ \\
Computer science & $2$ $(0.17)$ & $0.50$ $(0.00)$ & $0.50$ $(0.00)$ \\
\noalign{\smallskip}
\textit{Digg} user reply & $2$ $(0.25)$ & $0.10$ $(0.20)$ & $0.05$ $(0.28)$ \\
Emails at \textit{Enron}	& $2$ $(0.00)$ & $0.01$ $(0.57)$ & $0.50$ $(0.00)$ \\
\textit{Facebook} wall post & $2$ $(0.00)$ & $0.10$ $(0.18)$ & $0.01$ $(0.40)$ \\
Emails at EU res. inst. & $2$ $(0.00)$ & $0.50$ $(0.00)$ & $0.15$ $(0.60)$ \\
\noalign{\smallskip}
\textit{Amazon} products $1$ & $4$ $(0.33)$ & $0.50$ $(0.00)$ & $0.50$ $(0.00)$ \\
\textit{Amazon} products $2$ & $2$ $(0.00)$ & $0.50$ $(0.00)$ & $0.50$ $(0.02)$ \\
\noalign{\smallskip}
\textit{Flickr} images metadata & $3$ $(0.33)$ & $0.01$ $(0.80)$ & $0.25$ $(0.37)$ \\
\noalign{\smallskip}
\textit{Oregon} aut. systems & $2$ $(0.17)$ & $0.40$ $(0.23)$ & $0.01$ $(0.93)$ \\
\textit{Gnutella} file sharing $1$ & $5$ $(0.58)$ & $0.15$ $(0.34)$ & $0.15$ $(0.34)$ \\
\textit{Gnutella} file sharing $2$& $5$ $(0.42)$ & $0.15$ $(0.34)$ & $0.10$ $(0.36)$ \\
\noalign{\smallskip}
\textit{Foldoc} dictionary	&$2$ $(0.17)$ & $0.50$ $(0.00)$ & $0.50$ $(0.02)$ \\
\noalign{\smallskip}
\textit{Wikipedia} votes	& $2$ $(0.00)$ & $0.01$ $(0.26)$ & $0.01$ $(0.76)$ \\
\textit{Brightkite} friendship & $2$ $(0.00)$ & $0.05$ $(0.37)$ & $0.05$ $(0.83)$ \\
\textit{Epinions} trust & $2$ $(0.00)$ & $0.01$ $(0.94)$ & $0.01$ $(0.70)$ \\
\textit{Slashdot} friendship & $2$ $(0.17)$ & $0.01$ $(0.30)$ & $0.01$ $(0.58)$ \\
\textit{Wikipedia} interactions& $4$ $(0.47)$ & $0.01$ $(0.42)$ & $0.05$ $(0.44)$ \\
\textit{Gowalla} friendship & $2$ $(0.00)$ & $0.05$ $(0.33)$ & $0.05$ $(0.03)$ \\
\noalign{\smallskip}
Broad-topic queries	& $4$ $(0.33)$ & $0.10$ $(0.26)$ & $0.15$ $(0.34)$ \\
\textit{google.com} internal & $2$ $(0.00)$ & $0.15$ $(0.34)$ & $0.15$ $(0.36)$ \\
\textit{nd.edu} domain & $4$ $(0.08)$ & $0.01$ $(0.48)$ & $0.50$ $(0.32)$ \\
\textit{Baidu} articles & $2$ $(0.00)$ & $0.10$ $(0.22)$ & $0.05$ $(0.50)$ \\
\noalign{\smallskip}\hline
\end{tabular}
\end{table}

\subsubsection{Analysis of the sampling methods}

The analysis of the sampling methods reveals a high level of diversity
in their effectiveness (Fig.~\ref{fig:bestsizesampl} and Table~\ref{tbl:resS1}). Fig.~\ref{subfig:freqsampl}
shows that under simplification methods RN and RL, local properties
are best preserved for the largest size of the simplified networks
($s=0.5$). In contrast, RD and BF perform best for smaller sizes,
between $s=0.01$ and $s=0.15$, for the majority of the networks
(i.e., the local minimum of $A$ is around these values for most of
the networks).

Fig.~\ref{subfig:localsampl} and~\ref{subfig:globsampl} shows
the average $A$ over all networks for the local and global properties,
respectively. For the former, all methods behave in a similar manner.
In particular, the best fit of local properties is reached for larger
simplified networks; still, RD and BF show some deviation, indicating
that for several networks smaller sizes also provide good fits. For
the global properties, RN and RL show similar behavior again because
the best preservation is achieved on smaller simplified networks ($s=0.15$).
For BF and RD, the local and global minima are reached for larger
simplified networks.

Table~\ref{tbl:sum} shows the best sizes of simplified networks
for the preservation of each network property. RN and RL perform similarly
because both provide better preservation of local properties for the
largest simplified networks. On the other hand, for RD, degree is
best preserved for smaller networks, whereas for medium-sized networks,
out-degree and clustering are best preserved. For BF, distributions
of degree, out-degree and in-degree change the least for $s=0.01,0.15$.
However, the methods behave in a different manner when preserving
global properties. Only RN preserves density and degree mixing well
on smaller simplified networks, whereas RD, BF and RL work best for
$s=0.5$.

Finally, we analyze how the preservation of local properties depends
on the size and type of the original networks (Table~\ref{tbl:resS1}).
We omit the results for RN and RL because in all cases except two,
the best size is $s=0.5$. In contrast, the effectiveness of RD is
partially correlated to the original network size because medium-sized
networks ($n=50000-200000$) are best preserved for smaller simplified
network sizes ($s=0.01-0.1$), whereas large networks ($n=200000-500000$)
are best preserved for larger values of $s$. However, as indicated
by the dependence on network type, the local properties of on-line
social networks and Web graphs are best preserved for smaller sizes
$s=0.01-0.15$, whereas the local properties of citation and co-purchase
networks are best preserved for $s=0.25-0.35$. All differences are
statistically significant ($p<0.05$, one-way ANOVA), which rejects
the null hypothesis that there is no dependence between the effectiveness
of property preservation and network type. For both RD and BF, only
the properties of co-purchase and information networks are best preserved
for $s=0.5$. The results reveal no statistically significant influence
of network size or type on the performance of BF.

\subsubsection{Analysis of the merging methods}

\begin{figure}[t]
        \centering
        \subfigure[\label{subfig:bestsize}]{\includegraphics[width=0.47\columnwidth]{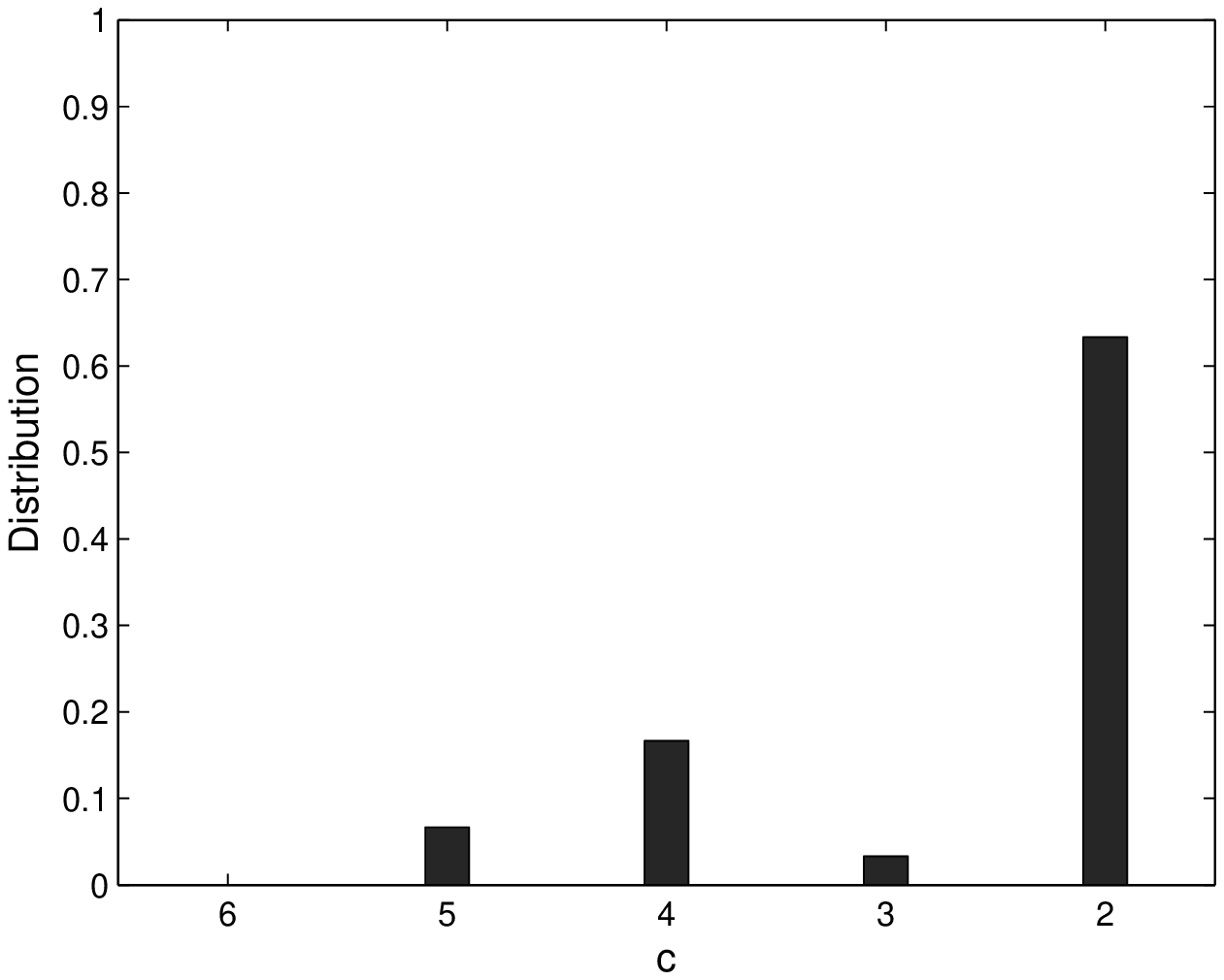}}\quad
				\subfigure[\label{subfig:avgdist}]{\includegraphics[width=0.47\columnwidth]{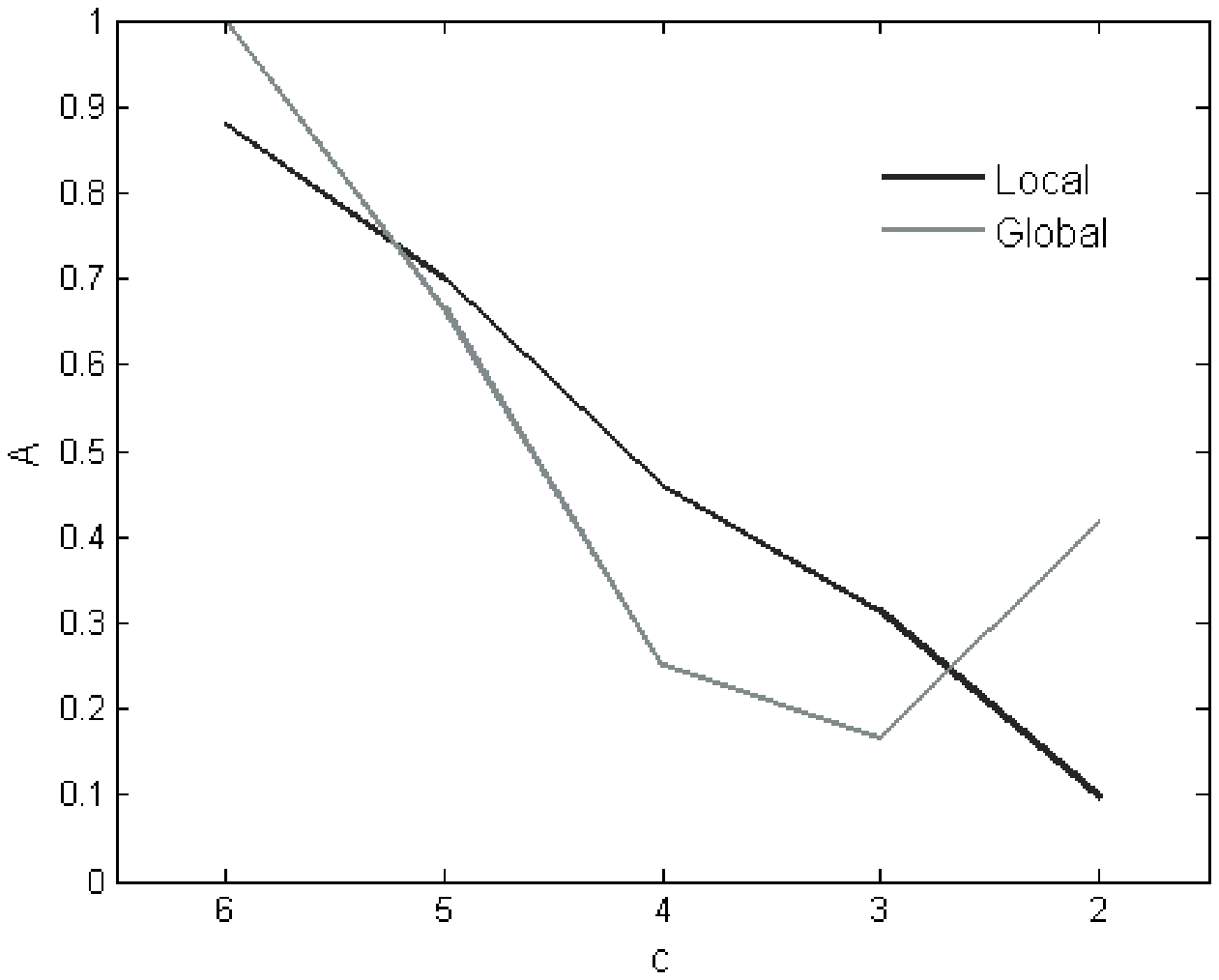}}
        \caption{The results for cluster-growing simplification. \subref{subfig:bestsize} Portion of networks 
				with the best size equal to $c$. \subref{subfig:avgdist} Distance between the original and simplified 
				networks ($A$ for the global and average $A$ over all networks for the local properties) as a function of $c$.}\label{fig:bestsizecg}
\end{figure}

The analysis of CG proves that the local network properties are best
preserved when $c=2$ for $22$ out of $30$ networks (Fig.~\ref{subfig:bestsize}
and Table~\ref{tbl:resS1}). Fig.~\ref{subfig:avgdist} shows the
average $A$ over all networks based on the local and global properties.
The local properties are best fitted for larger simplified networks
($c=2$), whereas for $c=3,4$ the simplified networks best fit the
global properties of the original networks.

\begin{table}[t]
\scriptsize
\centering
\caption{\label{tbl:sum}The best sizes $c$ or $s$ for the preservation of local network properties with corresponding $A$, and $\rho$ for the global properties.}
\begin{tabular}{lccccc}
\noalign{\smallskip}
Property & CG & RN & RD & BF & RL  \\\noalign{\smallskip}\hline\noalign{\smallskip}
Degree & $2$ $(0.04)$ & $0.50$ $(0.00)$ & $0.15$ $(0.52)$ & $0.15$ $(0.61)$ & $0.50$ $(0.00)$ \\
In degree & - & $0.50$ $(0.00)$ & $0.50$ $(0.26)$ & $0.01$ $(0.75)$ & $0.50$ $(0.02)$ \\
Out degree & - & $0.50$ $(0.00)$ & $0.30$ $(0.43)$ & $0.01$ $(0.61)$ & $0.50$ $(0.01)$ \\
Clustering & $2$ $(0.25)$ & $0.50$ $(0.00)$ & $0.25$ $(0.44)$ & $0.50$ $(0.10)$ & $0.50$ $(0.00)$ \\
Betweenness & $2$ $(0.00)$ & $0.50$ $(0.00)$ & $0.50$ $(0.08)$ & $0.50$ $(0.05)$  & $0.50$ $(0.01)$ \\
\noalign{\smallskip}
Density & $2$ $(0.89)$ & $0.10$ $(0.97)$ & $0.45$ $(0.95)$ & $0.50$ $(0.95)$ & $0.50$ $(0.91)$ \\
Degree mixing & $3$ $(0.34)$ & $0.35$ $(0.66)$ & $0.50$ $(0.77)$ & $0.50$ $(0.97)$ & $0.50$ $(0.63)$ \\
Transitivity & $4$ $(0.36)$ & $0.50$ $(0.99)$ & $0.50$ $(0.99)$ & $0.50$ $(0.99)$ & $0.50$ $(0.83)$ \\
\noalign{\smallskip}
\hline
\end{tabular}
\end{table}

Table~\ref{tbl:sum} shows the results obtained for the preservation
of each property. Most of the properties are best preserved for larger
simplified networks ($c=2$), with the exception of degree mixing
and transitivity, where $c=5$ and $c=6$, respectively.

The best size for preserving local network properties (Table~\ref{tbl:resS1})
does not depend on the original network size or type (i.e., the differences
in property preservation, which would depend on the size and type
of the original networks, are not statistically significant). Still,
if we divide the networks roughly by type, i.e., information, social
and technological, the correlation between the type and the effectiveness
becomes statistically significant (i.e., the null hypothesis that
there are no differences in property preservation, which would depend
on network type, is rejected, with $p<0.05$, one-way ANOVA). Thus,
the local properties of social networks are best preserved for $c=2$,
in contrast to the case of technological networks, for which $c>2$.

\subsubsection{Discussion}

The findings of the first part of the study confirm the negative correlation
between the size of the simplified networks and their similarity to
the original networks because larger simplified networks are more
similar to the original ones in most cases. The latter has also been
proved by other studies, for example,~\cite{LKJ06}. RD and BF are
more effective for smaller simplified networks, which is consistent
with the findings of other authors. Particularly, Doerr and Blenn~\cite{DB13}
revealed a solid estimate of an original network for $s=0.2-0.3$
and $s=0.1$ in the case of preserving average node degree and the
power-law degree exponent, respectively. In addition, Leskovec and
Faloutsos~\cite{LF06} obtained a good fit for original networks
under several sampling methods for $s=0.15$. Thus, our results advance
those reported in these studies and reveal distinctions in the extent
of property preservation among different types and sizes of networks,
which are the most obvious for RD.

\subsection{\label{subsec:compare}Comparison of the effectiveness of the simplification
methods}

In the second part of our study, we compare the performance of different
simplification methods. We focus on size $c=2$ for CG and $s=0.1$
for sampling methods for two reasons. First, we select $s=0.1$ as
the middle size among the best sizes determined in the first part
of the study. Second, $s=0.1$ is suitable for the comparison of BP
and CG, for which the mean sizes of simplified networks are $s=0.03$
and $s=0.12$, respectively.

\begin{table}[t]
\scriptsize
\centering
\caption{\label{tbl:resall}The best, second-best and worst methods for the preservation of local 
network properties with corresponding $A$, and $\rho$ for the global properties.}
\begin{tabular}{lccc}
\noalign{\smallskip}
Property & Best  & Second-best & Worst\\\noalign{\smallskip}\hline\noalign{\smallskip}
Degree & BF $(0.25)$ & RD $(0.26)$ & RL $(0.84)$ \\
In degree &  RD/BF $(0.26)$ & RL $(0.70)$ & RN $(0.77)$  \\
Out degree &  RD $(0.32)$ & BF $(0.33)$ & RL $(0.70)$  \\
Clustering & RD $(0.30)$ & BF $(0.35)$ & RL $(0.81)$  \\
Betweenness & BF $(0.21)$ & RD $(0.27)$ & BP $(0.75)$  \\
\noalign{\smallskip}
Density & RN $(0.96)$ & BF $(0.91)$ & BP $(0.76)$ \\
Degree mixing & BF $(0.92)$ & RN $(0.62)$ & BP $(0.21)$ \\
Transitivity & RN $(0.94)$ & RD $(0.92)$ & CG $(0.22)$ \\
\noalign{\smallskip}
\hline
\end{tabular}
\end{table}

\begin{table}[t]
\scriptsize
\centering
\caption{\label{tbl:resS2}The best, second-best and worst methods for preserving local properties 
of networks with corresponding $A$.}
\begin{tabular}{lccc}
\noalign{\smallskip}
Network & Best & Second-best & Worst \\\noalign{\smallskip}\hline\noalign{\smallskip}
\textit{High E. Particle Phys.} & RD $(0.10)$ & BF $(0.17)$ & RL $(0.97)$  \\
\textit{High E. Phys.} &  BF $(0.07)$ & RD $(0.20)$ & RL $(0.96)$  \\
\textit{NBER US} patents &	BF $(0.07)$ & BP $(0.13)$ & RN/RL $(0.80)$  \\
\textit{Citeseer} publications &	 RD $(0.07)$ & BF $(0.20)$ & RL $(0.93)$  \\
\noalign{\smallskip}
\textit{PGP} web-of-trust	&	CG $(0.13)$ & BP $(0.20)$ & RN $(0.93)$ \\
\textit{High E. Phys.} archive &  RD $(0.07)$ & BF $(0.20)$ & RL $(0.93)$  \\
\textit{Astro Phys.} archive & RD $(0.07)$ & BF $(0.13)$ & RL $(1.00)$ \\
\textit{Cond. Matters} archive &	 BF $(0.00)$ & RD $(0.20)$ & RL $(1.00)$  \\
Computer science & BF $(0.07)$ & RD $(0.27)$ & RL $(1.00)$  \\
\noalign{\smallskip}
\textit{Digg} user reply & RD $(0.17)$ & CG/BP $(0.33)$ & RL/RN $(0.60)$  \\
Emails at \textit{Enron} &	 BP $(0.27)$ & RD $(0.33)$ & RN $(0.73)$ \\
\textit{Facebook} wall post &	RD $(0.07)$ & BP $(0.17)$ & RL $(1.00)$  \\
Emails at EU res. inst. &	RL $(0.13)$ & BP $(0.20)$ & RD $(0.73)$  \\
\noalign{\smallskip}
\textit{Amazon} products $1$ &  BF $(0.00)$ & BP $(0.27)$ & RL $(1.00)$  \\
\textit{Amazon} products $2$ & BF $(0.03)$ & CG $(0.10)$ & RL $(1.00)$ \\
\noalign{\smallskip}
\textit{Flickr} images metadata &	 RD/BF $(0.33)$ & RN $(0.47)$ & RL $(0.73)$  \\
\noalign{\smallskip}
\textit{Oregon} aut. systems &	RD $(0.07)$ & BP $(0.20)$ & RN $(0.80)$  \\
\textit{Gnutella} file sharing $1$	&	 BF $(0.13)$ & BP $(0.30)$ & RL $(0.70)$  \\
\textit{Gnutella} file sharing $2$	&	BF $(0.13)$ & BP $(0.30)$ & RL $(0.70)$  \\
\noalign{\smallskip}
\textit{Foldoc} dictionary	& BF $(0.03)$ & CG/BP $(0.13)$ & RL $(1.00)$  \\
\noalign{\smallskip}
\textit{Wikipedia} votes	&	BP $(0.13)$ & RN $(0.27)$ & BF $(0.60)$ \\
\textit{Brightkite} friendship &	 RD $(0.13)$ & BP $(0.20)$ & RL $(0.93)$  \\
\textit{Epinions} trust & BP $(0.03)$ & RL $(0.17)$ & BF $(0.87)$   \\
\textit{Slashdot} friendship &	BP $(0.07)$ & RD $(0.23)$ & RL $(0.83)$  \\
\textit{Wikipedia} interactions &	 BP $(0.07)$ & BF $(0.33)$ & RN $(0.40)$  \\
\textit{Gowalla} friendship &	 RD $(0.07)$ & BP $(0.27)$ & RL $(1.00)$ \\
\noalign{\smallskip}
Broad-topic queries	&  RD $(0.07)$ & BP $(0.27)$ & RN $(0.73)$ \\
\textit{google.com} internal &	 RD $(0.10)$ & BF $(0.17)$ & RL $(0.93)$  \\
\textit{nd.edu} domain &	CG $(0.07)$ & BP/BF $(0.13)$ & RL/RN $(0.80)$ \\
\textit{Baidu} articles & RD $(0.00)$ & BP $(0.27)$ & RL $(0.90)$  \\
\noalign{\smallskip}
\hline
\end{tabular}
\end{table}

\subsubsection{Analysis}

\begin{figure}[t]
        \centering
        \subfigure[\label{subfig:dens}]{\includegraphics[width=0.47\columnwidth]{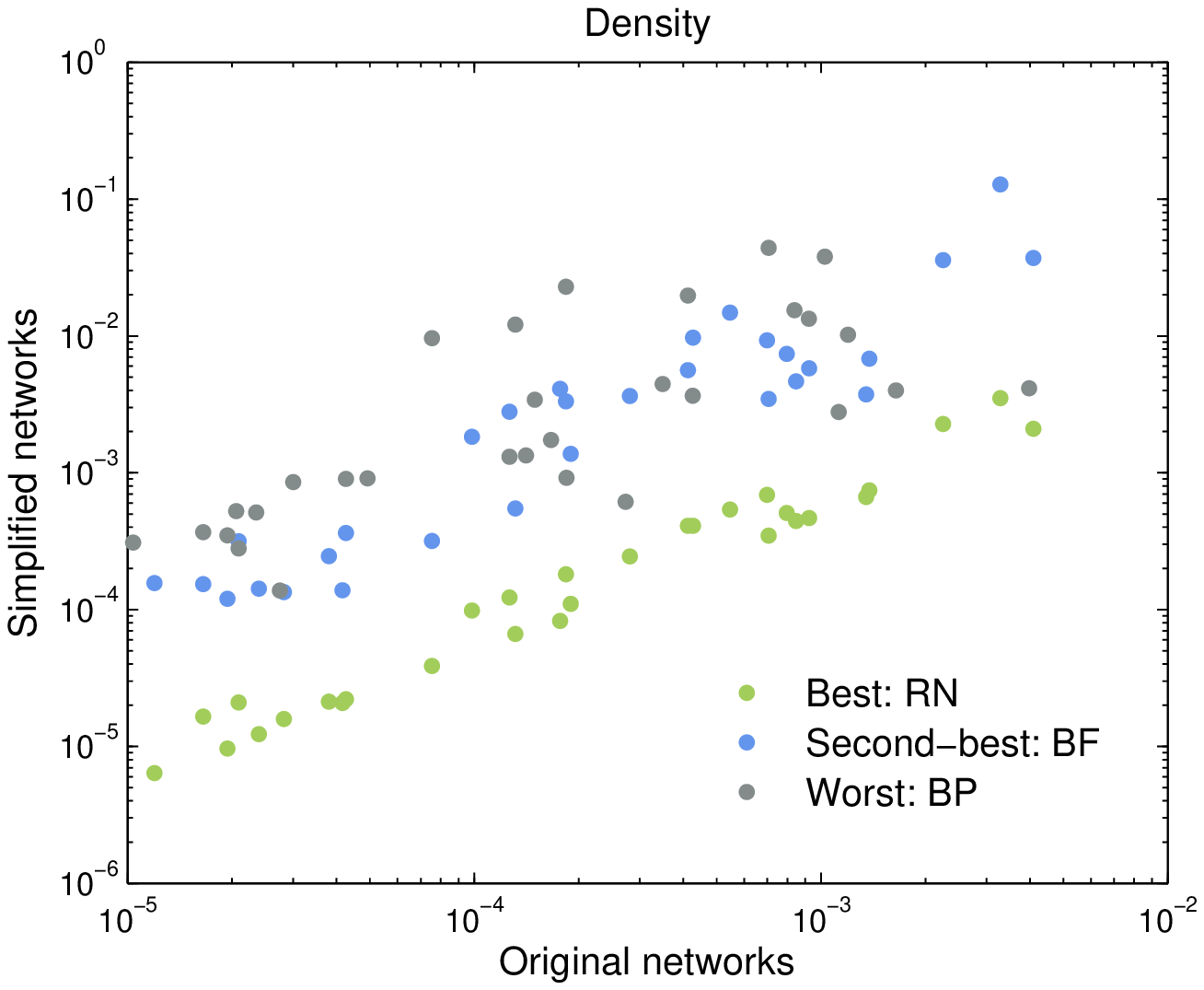}}\quad
				\subfigure[\label{subfig:assort}]{\includegraphics[width=0.47\columnwidth]{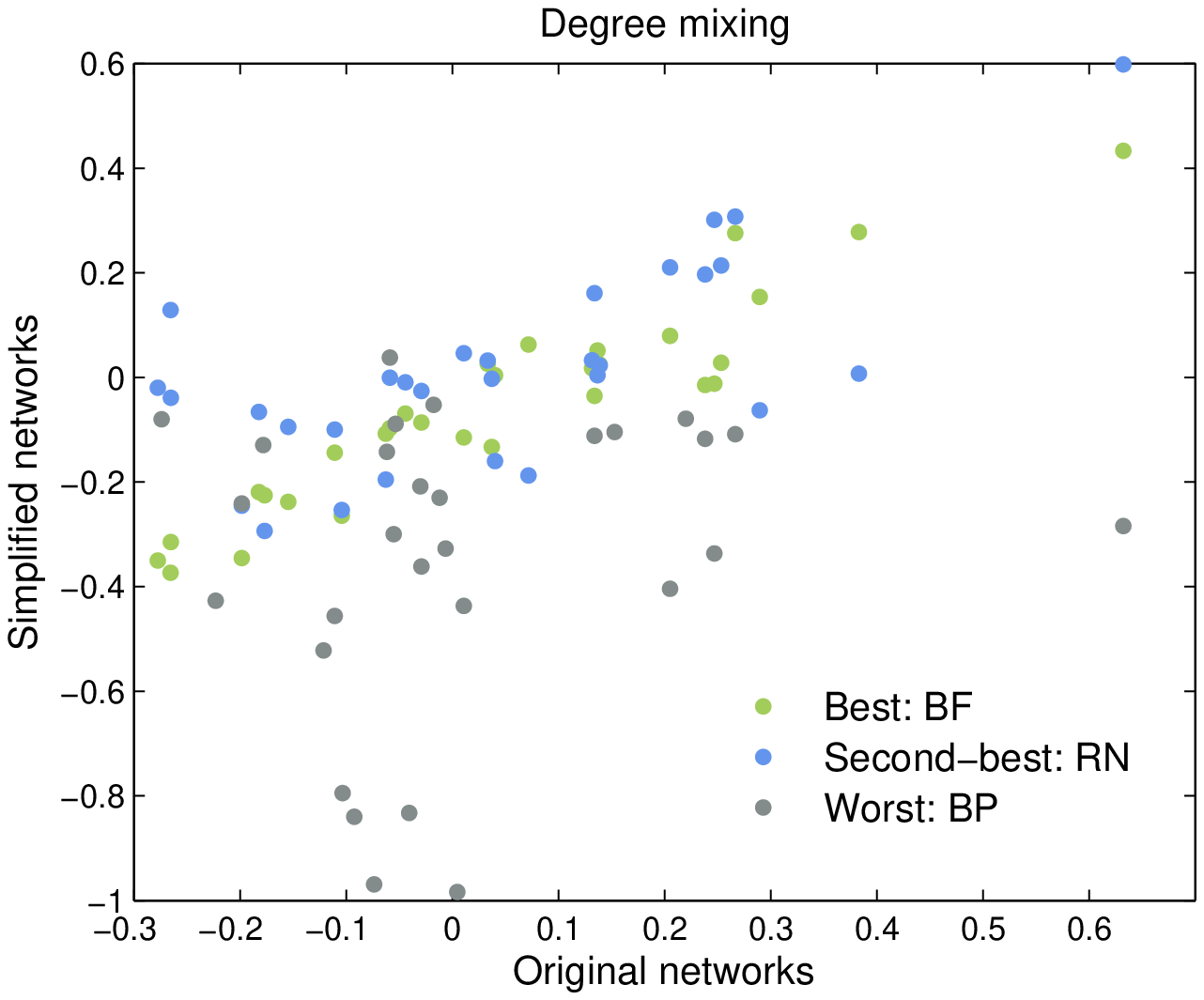}}\\
				\subfigure[\label{subfig:trans}]{\includegraphics[width=0.47\columnwidth]{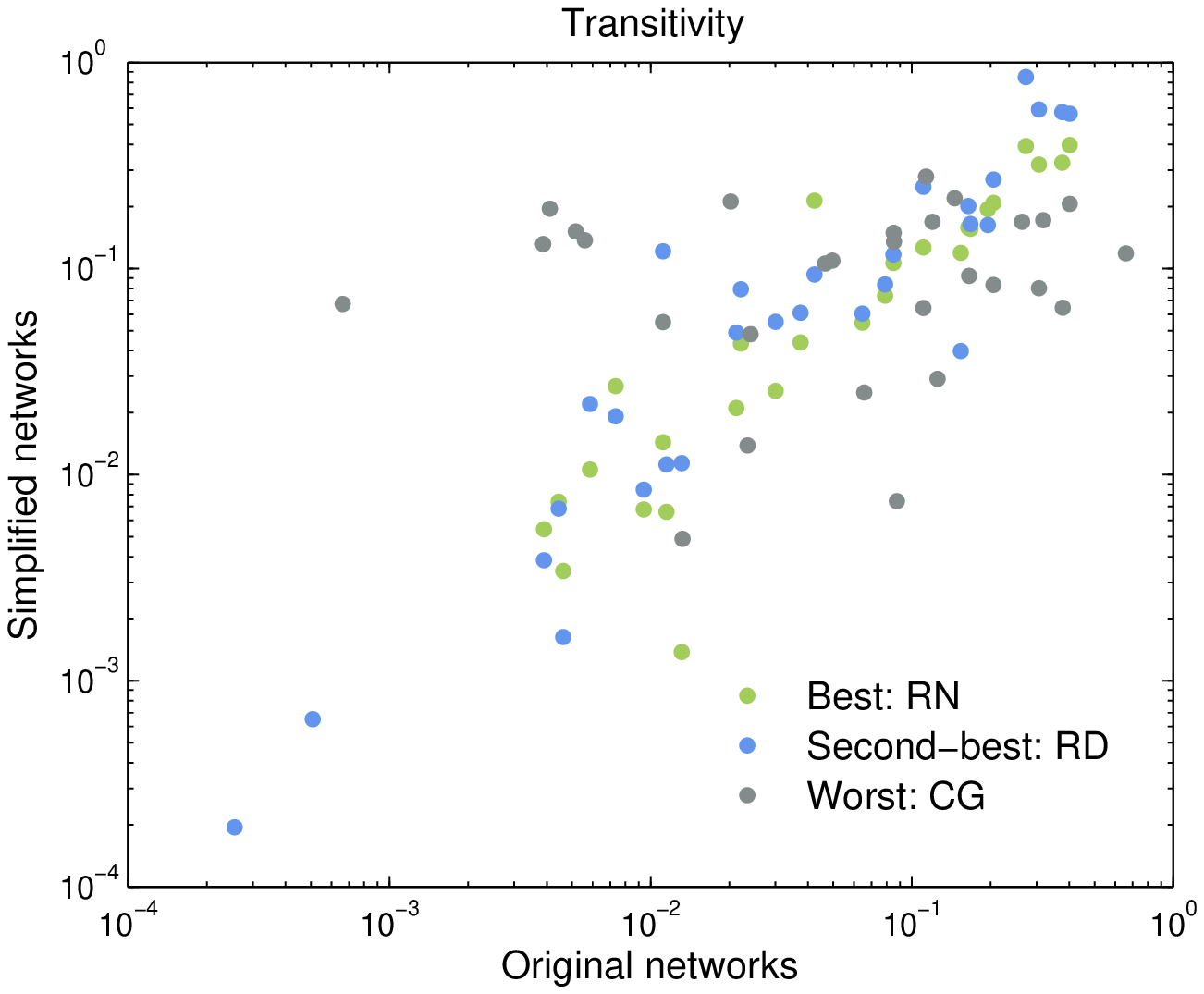}}
        \caption{Relationship between the global properties of the original and the simplified networks 
				for the best, second-best and worst method. \subref{subfig:dens} Density. \subref{subfig:assort} 
				Degree mixing. \subref{subfig:trans} Transitivity.}\label{fig:global}
\end{figure}

\begin{figure*}[p]
        \centering
        \subfigure[\label{subfig:deg}]{\includegraphics[width=0.47\columnwidth]{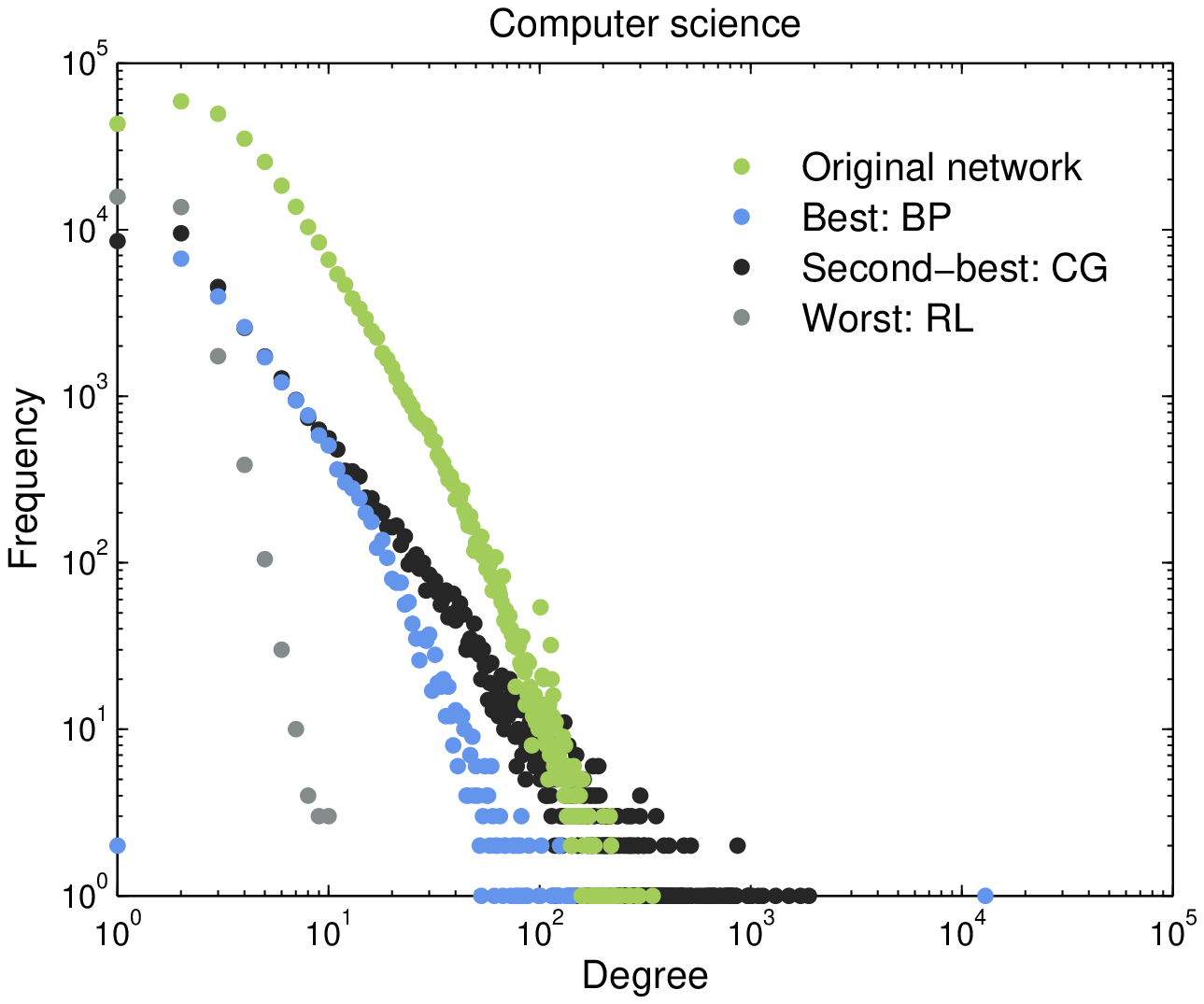}}\quad
				\subfigure[\label{subfig:indeg}]{\includegraphics[width=0.47\columnwidth]{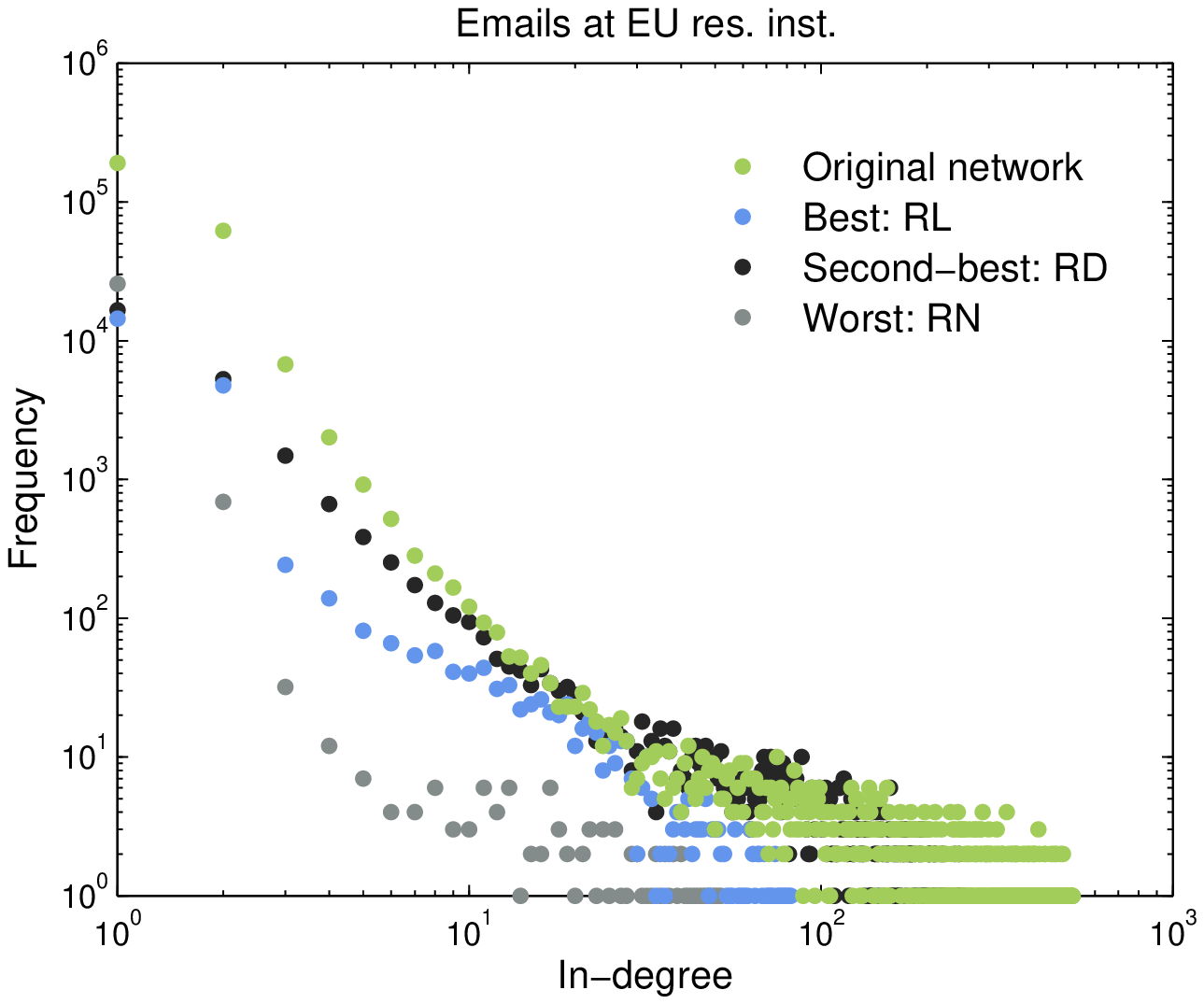}}\\
				\subfigure[\label{subfig:outdeg}]{\includegraphics[width=0.47\columnwidth]{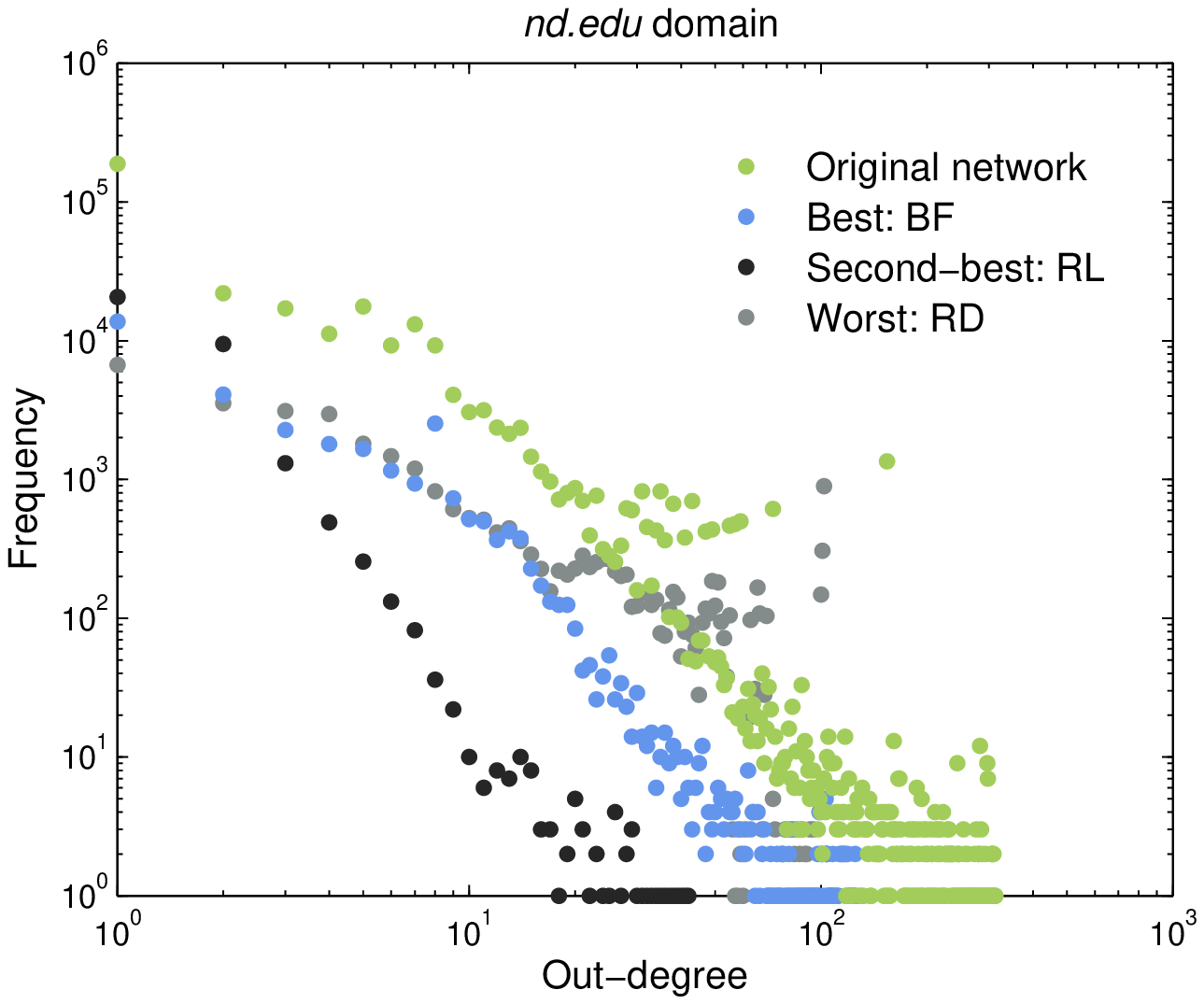}}\\
				\subfigure[\label{subfig:clus}]{\includegraphics[width=0.47\columnwidth]{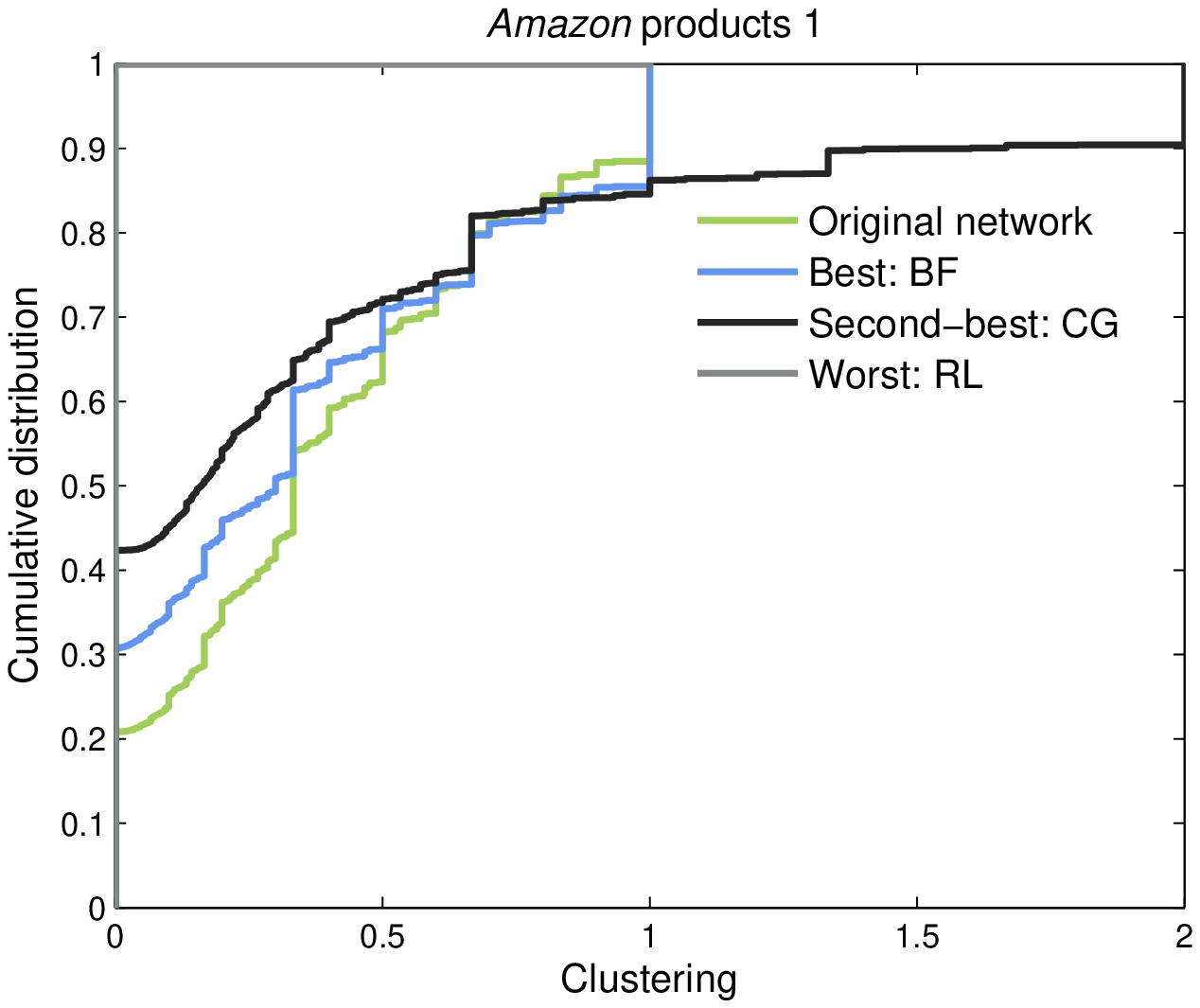}}\quad
				\subfigure[\label{subfig:betw}]{\includegraphics[width=0.47\columnwidth]{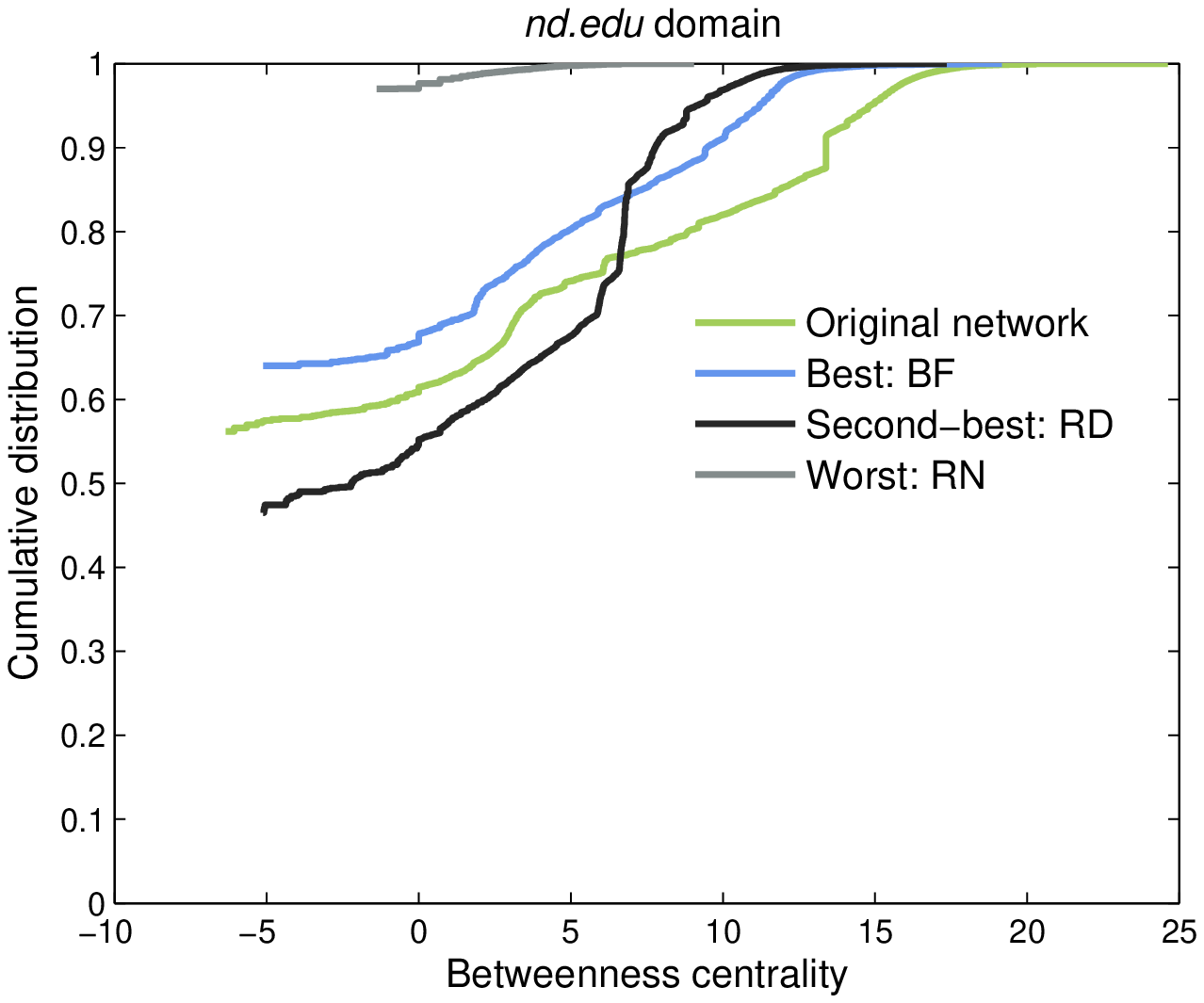}}
        \caption{Examples of comparison of the local properties for the original and simplified networks with 
				the best, second-best and worst methods. \subref{subfig:deg} Degree distribution. \subref{subfig:indeg} 
				In-degree distribution. \subref{subfig:outdeg} Out-degree distribution. \subref{subfig:clus} Cumulative 
				distribution of clustering. \subref{subfig:betw} Cumulative distribution of of betweenness centrality.}\label{fig:local}
\end{figure*}

First, we determine the best method for preserving a specific property
(Table~\ref{tbl:resall}). Global properties are best preserved under
RN and BF, whereas merging methods provide the worst preservation.
Fig.~\ref{fig:global} compares the best, second-best and worst methods
with respect to all global properties. For local properties, BF and
RD perform the best, particularly BF for the degree and betweenness
centrality, whereas RD performs best for the out-degree and clustering.
However, RL proves to be the worst method because it preserves the
degree, out-degree and clustering to the lowest extent. Examples of
local property preservation for the analyzed networks are presented
in Fig.~\ref{fig:local}.

For a complete assessment of the effectiveness of the simplification
methods, we compare the performance of the methods for each network
based on the preservation of local properties. Results are represented
in Table~\ref{tbl:resS2}. For $23$ networks, the best methods are
RD and BF. The analysis reveals a dependence between network type
and method effectiveness because BP performs the best for on-line
social networks and BF performs the best for Internet and co-purchase
networks. The differences among the network types are statistically
significant ($p<0.05$, one-way ANOVA). For the second-best methods,
the distinctions are less evident. Still, BP proves to be effective
for other types of networks (Internet, communication networks, Web
graphs). The worst method for preserving local properties is RL (for
$22$ networks), followed by RN (for $8$ networks). On the other
hand, BF is the worst with respect to only two on-line social networks.
The results also prove the statistically significant dependence (i.e.,
reject the null hypothesis that there are no dependencies between
the network size and the effectiveness of the simplification methods,
with $p<0.05$, one-way ANOVA) between the worst method and network
size. For smaller networks ($n<50000$), the worst method for preserving
local properties is RL, whereas for larger ones, the worst method
is RN.

\subsubsection{Discussion}

The results of the second part of the study reveal several distinctions
in the behavior of the simplification methods. RD and BF proved the
best for preserving the local properties of networks, whereas for
global properties, RN outperforms the other methods. However, RL and
merging methods show the worst performance. These findings are consistent
with the results of the study reported in~\cite{LF06}, where RD
had a better performance than RN and RL (other methods are not considered
in the aforementioned study).

In addition to comparing the methods for $s=0.1$, we also compare
them for larger simplified networks ($s=0.5$). The results are not
presented because there are no significant changes in the results
(i.e., the same methods are the best and the worst for $s=0.1$).

In addition, we observe how the size of the largest weakly connected
component (LWCC) changes under simplification to explain the differences
in the methods' performance. The LWCC of the original networks, on
average, consists of $59\%$ of all nodes. The size of the LWCC of
the simplified networks under all methods depends strongly on the
simplified network size (i.e., the size of the LWCC of the smallest
simplified networks is the smallest). However, RN and RL show similar
performance because the sizes of the LWCC for both methods vary from
$1\%$ for $s=0.1$ to $40\%$ for $s=0.5$. Still, RL produces the
most disconnected components. In contrast, simplification via RD and
BF produces networks with a clearly larger LWCC because the sizes
vary from $25\%$ for $s=0.01$ to $60\%$ for $s=0.5$. Therefore,
networks simplified by RD and BF feature a larger LWCC and smaller
number of components, which is more similar to the characteristics
of the original networks. Based on this finding, the predominance
of RD and BF over RN and RL can be confirmed.


\section{\label{sec:conc}Conclusions}

Network simplification is an adequate tool for studying large networks
for several reasons. In addition to the obvious advantages, including
faster analysis and more efficient visualization, the simplification
can significantly improve the understanding of large networks. For
example, data regarding the systems described by networks can often
be missing or incomplete, and thus, networks can be considered a sampled
variety of the original systems (e.g., identifying Internet map~\cite{CM05,VBDAZK07}).
For this reason, understanding how similar the original and sampled
system are is essential.

This study addressed three aspects of real-world network simplification.
First, we focused on a comparison of original and simplified networks.
Second, we determined what size of simplified network most adequately
fits the properties of the original networks. Finally, we compared
the effectiveness of several simplification methods. We analyzed six
simplification methods with respect to $30$ real-world networks and
compared the simplified and original networks based on several properties,
including degree, in-degree, out-degree and betweenness centrality
distribution, clustering coefficient, density, degree mixing and transitivity.

The results show that the goodness of property preservation depends
on the size of the simplified networks. Larger simplified networks
fit original networks better; nevertheless, properties are adequately
preserved for smaller sizes close to $10\%$ the size of the original
networks, especially for random node selection based on degree and
breadth-first sampling. Thus, the decision regarding how small a simplified
network should be depends on the size of the original network and
the purpose of the simplified network. If we can simplify a network
by $50\%$, we can provide for the best fit of the original network
properties. However, if the network is large, $50\%$ of the original
size is not a sufficient reduction. In that case, $10\%$ of the original
network size allows for the adequate preservation of important properties.
Furthermore, the findings of this study reveal that random node selection
based on degree and breadth-first sampling are the best methods, whereas
merging methods performed the worst.

Future work will mainly focus on other characteristics that affect
the effectiveness of the simplification process. Moreover, instead
of focusing solely on similarities, we will analyze typical distinctions
between original and simplified networks. Furthermore, other ways for 
comparing simplified networks with original for their similarity could 
also be considered, for example comparing the backbones of networks~\cite{SBV09},
their community structure~\cite{OFRPMFJ12} or density of edges in 
subnetworks~\cite{GF14}. Based on this and future
studies, a wide range of principles underlying the simplification
of real-world networks could be extracted. The application of such
principles should allow for the determination of the most suitable
simplification method for specific networks, which would allow for
more efficient simplification and a better understanding of large
real-world networks.


\section*{Acknowledgment}

The work has been supported by the Slovene Research Agency \textit{ARRS}
within the research program P2-0359.


\bibliographystyle{elsarticle-num}


\end{document}